\documentclass[aip,reprint,twocolumn,letterpaper]{revtex4-2}
\bibstyle{aipnum4-2}
\usepackage{amsmath}
\usepackage{graphicx}
\usepackage[labelformat=empty]{subcaption}
\usepackage{xcolor}
\usepackage[bookmarks=false,         
     pdfnewwindow=true,      
     colorlinks=true,    
     linkcolor=blue,     
     citecolor=blue,     
     filecolor=blue,  
     urlcolor=blue,      
final=true
]{hyperref}
\usepackage{mathrsfs} 
\usepackage[font=small,labelfont=bf,labelsep=period]{caption}
\newcounter{subfig}[figure]


\begin{document}

\title{Design, analysis, and experimental validation of a stepped plate parametric array loudspeaker}

\author{Woongji Kim}
\author{Beomseok Oh}
\author{Chayeong Kim}
\altaffiliation{Current address: Mechatronics Research, 
	Samsung Electronics,
	Hwasung,
	Gyeonggi 18448,
	Korea}
\author{Wonkyu Moon}
\email{wkmoon@postech.ac.kr}
\affiliation{Department of Mechanical Engineering, 
	Pohang University of Science and Technology,
	Pohang,
	Gyeongbuk 37673,
	Korea}

\date{April 23, 2025}

\begin{abstract}
This study investigates the design and analysis of a stepped plate parametric array loudspeaker (SPPAL) as an alternative to conventional array-based parametric loudspeakers. The SPPAL utilizes a single Langevin-type ultrasonic transducer coupled with a flexural stepped plate to generate narrow-beam audible sound via nonlinear acoustic interaction. To evaluate and optimize the performance of the SPPAL, an integrated modeling framework is developed, consisting of an approximate analytical 3D model for transducer dynamics, an equivalence ratio formulation to relate stepped plate and rigid piston behavior, and a spherical wave expansion method for nonlinear sound field simulation. The dual-resonance behavior of the transducer is optimized through multi-objective analysis to enhance low-frequency audio performance. Experimental validation includes frequency response and modal analysis of the transducer, as well as sound field measurements. The analytical methods are further verified through comparison with experimental data. Furthermore, combination resonance--an unintended structural excitation resulting from intermodulation--is identified as an inherent phenomenon in SPPAL operation. The findings offer practical guidance for the development of efficient, compact, and manufacturable parametric array loudspeakers employing plate-based flexural vibration.
\end{abstract}

\maketitle

\section{\label{sec:1} Introduction}

Parametric array loudspeakers (PALs) utilize the nonlinear interaction of high-frequency ultrasonic waves in air to produce audible sound through self-demodulation. Since Westervelt's seminal work,\cite{Westervelt1963} this principle has offered a fundamentally different approach to sound reproduction--enabling audio beams of exceptional directivity from apertures far smaller than those required by conventional loudspeakers. Such capabilities have positioned PA technology as an attractive solution for targeted audio delivery in applications ranging from museum exhibits and digital signage to personal listening zones and quiet zones.\cite{Yoneyama1983,Pompei1999,Gan2012,Zhong2022a,Zhong2025} Nonetheless, realizing sufficient sound pressure levels for practical use remains a persistent challenge, owing to the intrinsically low efficiency of nonlinear conversion,\cite{Li2023} thereby motivating continued innovation in transducer design and system architecture.
\par
Traditional PAL configurations employ arrays of single-resonance ultrasonic transducers, which offer practical advantages such as flexible array geometries and the ability to implement advanced control techniques like beam steering or focusing. Commercially available emitters such as Murata's MA40S4S are widely used due to their stable phase response and compact form factor.\cite{Zhong2024} In addition, micro-machined ultrasonic transducers (MUTs), including capacitive (CMUTs)\cite{Wygant2009} and piezoelectric (PMUTs) types,\cite{Lee2009,Je2015} have also been actively investigated for PAL applications to enable miniaturized, high-frequency arrays with improved integration capability. However, these array-based systems are often constrained by increased system complexity, phase-matching requirements among multiple elements, and high implementation costs, especially when precise phase and amplitude matching is required to preserve beam quality.
\par
To address these limitations, a promising alternative has emerged in the form of the stepped plate parametric array loudspeaker (SPPAL),\cite{Je2010} which eliminates the need for complex transducer arrays by employing a single ultrasonic transducer. Initially developed to achieve phase-compensated ultrasonic radiation,\cite{Barone1972} the SPPAL incorporates annular steps into a flat plate (FP), allowing it to approximate the coherent output of a rigid piston (RP) radiator. This phase compensation mechanism enables spatially coherent wavefronts despite the use of a flexural structure. By combining this simplified architecture with a single Langevin-type transducer, the SPPAL offers several advantages, including reduced system complexity, a significantly enlarged aperture, and lower implementation cost. These advantages position the SPPAL as a compelling solution to the practical challenges that have thus far limited the commercialization of array-based PAL systems. 
\par
Despite the promising advantages of the SPPAL concept, a systematic evaluation of its performance across a wide range of design parameters has not yet been fully explored. This is primarily due to the lack of analytical models that simultaneously offer both high fidelity and computational efficiency--features essential for navigating the complex design space of SPPALs.
\par
To begin with, accurate modeling of the ultrasonic transducer is critical. The SPPAL employs close dual-resonance (DR) operation to amplify both the carrier and sideband components, enhancing the low-frequency audio output of parametric arrays. However, the conventional 1D Langevin model\cite{RanzGuerra1975} lacks the ability to capture radial-longitudinal coupling and is thus insufficient for exploring extended design parameters. To overcome this limitation, this study adopts an approximate analytical three-dimensional (Approx. 3D) model that incorporates Poisson coupling,\cite{Iula1998} allowing for efficient DR prediction without full FEM. Similar modeling approaches have been reported in the literature, highlighting the significance of radial effects in piezoelectric transducers.\cite{Lin1995,Feng2006,Mancic2010} The present model builds on this foundation to enable broader parametric exploration. In particular, this model facilitates the evaluation of feasible horn-end radii for stepped horn structures. Extremely small radii lead to thin-neck geometries that are difficult and costly to fabricate, making such modeling essential for maintaining the SPPAL's cost-efficiency.
\par
Second, the stepped plate (SP) structure introduces significant variation in mechanical characteristics due to the addition of annular steps.\cite{Hwang2018} To mitigate this effect, the steps are constructed from a polymer material with lower density and Young's modulus than the base plate, thereby minimizing their influence on the plate's natural frequency and mode shape.\cite{Kim2022} Furthermore, due to the high modal sensitivity of higher-order modes in thin plates, classical plate theory (CPT) proves inadequate for accurate modeling. This work instead employs the modified Mindlin plate theory (MMPT) to obtain reliable initial estimates of the structural dimensions.\cite{Oh2023} Using this formulation, the plate geometry is designed with respect to the first local maximum distance, the nominal ultrasonic frequency, and the vibrational mode number, allowing global trends to be identified for a wide range of design conditions.
\par
Lastly, the prediction of parametric audio fields requires nonlinear acoustic modeling, which is computationally demanding. Traditional simulation tools, such as the time-domain KZK (Texas code)\cite{Lee1995} and frequency-domain KZK (Bergen code)\cite{Aanonsen1984,Tjotta2005} solvers, offer accurate results within the paraxial region, but are limited by high computational cost and directional constraints. In recent years, alternative approaches--such as the Gaussian Beam Expansion (GBE)\cite{Cervenka2013} and Spherical Wave Expansion (SWE)\cite{Zhong2020,Zhuang2025} methods derived from the Westervelt equation, as well as extended King integral formulations\cite{Li2024}--have emerged to overcome these limitations. In addition, numerical simulations based on the finite element method (FEM) have also been employed to model nonlinear acoustic propagation, particularly when complex boundary conditions must be considered.\cite{Kagawa1992,Campos1999,Cervenka2019} Moreover, Kuznetsov equation-based methods that incorporate local nonlinearity have shown promise for capturing complex acoustic phenomena.\cite{Zhong2021,Cervenka2022,Li2024} In this study, the SWE and FEM-based method are employed to conduct efficient and reliable simulations of the audible sound field generated by the SPPAL across a wide set of design parameters.
\par
This study presents a comprehensive evaluation of the SPPAL design framework by integrating accurate transducer modeling, mechanically robust SP design, and efficient nonlinear acoustic field analysis. Emphasis is placed on identifying key design parameters that enable DR operation and on validating the acoustic equivalence between SPs and RPs. By employing Approx. 3D transducer models and SWE-based acoustic simulations, the proposed approach enables rapid yet reliable exploration of a broad design space. Furthermore, a series of rigorous experiments were conducted to validate the proposed modeling methods and to investigate the practical behavior of the SPPAL. These measurements not only confirmed the predictive accuracy of the analysis framework, but also revealed the presence of combination resonance (CR)\cite{Nayfeh1995}--an inherent structural phenomenon not previously reported nor captured by existing models.
\par
The insights derived from this work not only deepen the understanding of structural-acoustic interactions in flexural ultrasonic radiators but also offer practical guidance for the implementation of high-performance PAL systems using compact and cost-effective transducer configurations.

\section{\label{sec:2} Parametric array loudspeakers with stepped plates}

This section examines the SP as a viable radiator for directional audio applications. The acoustic performance of the SP is compared against that of a RP, focusing on the propagation curve (PC) and beam pattern (BP) characteristics. The equivalence ratio (ER) is introduced to quantify the similarity between the SP and RP in terms of axial propagation. Finally, the role of DR transducer design is discussed to address the SPL drop-off at low audio frequencies.

\subsection{\label{subsec:21} Acoustic characterstics of rigid piston for parametric array loudspeakers}

\begin{figure}[ht]
    \begin{subfigure}[b]{8.5cm}
        \centering
        \includegraphics[width=\linewidth]{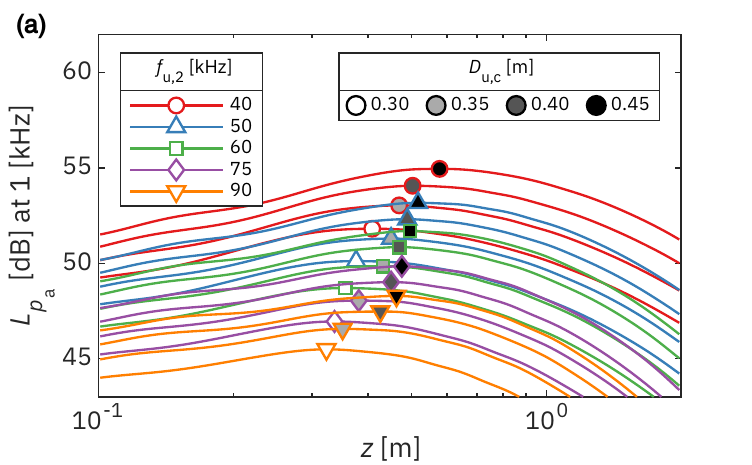}
        \caption{}
        \label{fig:1a}
    \end{subfigure}
    \begin{subfigure}[b]{8.5cm}
        \centering
        \includegraphics[width=\linewidth]{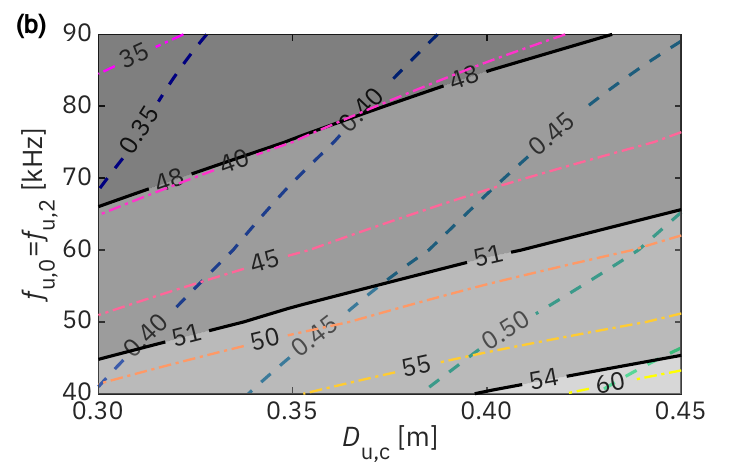}
        \caption{}
        \label{fig:1b}
    \end{subfigure}
    \caption{
        \label{fig:1}
        (a) PCs of the audio sound of $1$~kHz generated by circular RPs with $20$ combinations of design parameters, including the carrier frequencies, $f_{\mathrm{u,2}}$ and the ultra CD, $\mathscr{D}_\mathrm{u,c}$; (b) its summary contour plot. Solid lines: the critical audio SPL, $L_{p_\mathrm{a,c}}$; dashed lines: audio CD, $\mathscr{D}_{\mathrm{a,c}}$; dash-dotted lines: the aperture radii, $a$.
    }
\end{figure}

Since baffled circular RPs are the most well-known acoustic source even in PALs, the preliminary investigation begins with this type of radiator. In the following section, the circular SP is examined by leveraging its similarity to the circular RP to explore an optimal design strategy for the SPPAL at a given receiver location.
\par
For PAL with a baffled circular RP, the audio sound power is approximately proportional to $f_\mathrm{a}^{2.4}$, aperture size, $a^{2.6}$, and inversely proportional to ultrasound frequency, $f_\mathrm{u}^{-1.2}$ in the inverse-law far field.\cite{Zhong2023, Li2023} Similarly but, the following analysis focuses on the critical distance for audio sound (audio CD), $\mathscr{D}_\mathrm{a,c}$, defined as the location at which the on-axis audio pressure reaches its maximum as marked in Fig.~\ref{fig:1a}. The resulting relationship provides a guideline for selecting the carrier frequency and aperture radius to maximize the audio sound pressure at a given receiver location.
\par
First, the aperture radius, $a$ is determined based on the nominal ultrasound frequency, $f_\mathrm{u,0} = c_0/\lambda_\mathrm{u,0} = \{40, 50, 60, 75, 90\}$~kHz, corresponding to the first local maximum distance, $z_1 = a^2/\lambda_\mathrm{u,0} - \lambda_\mathrm{u,0}/4 = \{0.30, 0.35, 0.40, 0.45\}$~m, hereafter referred to as the critical distance of ultrasound (ultra CD), $\mathscr{D}_\mathrm{u,c}$. The parameter $f_\mathrm{u,0}$ is used either as the actual carrier frequency, $f_\mathrm{u,2}$, or as a reference design parameter for the SPPAL. This study adopts a finite set of design parameter combinations to ensure computational feasibility, as nonlinear acoustic effects--unlike linear ones--lack tractable analytical solutions and require numerically intensive simulations. To model the nonlinear field, the spherical wave expansion (SWE) method is employed, offering an efficient and rigorous solution to the Westervelt equation under the quasilinear approximation.\cite{Zhong2020} Atmospheric absorption is modeled according to ISO 9613-1,\cite{ISO9613} assuming a relative humidity of $70$\% and a temperature of $20$~\textcelsius. The surface normal velocities for the carrier and sideband components, $v_{\mathrm{u,2}}$ and $v_{\mathrm{u,1}}$, are each set to $0.1$~m/s. The modulation scheme used is lower sideband amplitude modulation (LSB-AM), with the audio frequency defined as $f_{\mathrm{a}} = f_{\mathrm{u,2}} - f_{\mathrm{u,1}}$. The rationale for this choice will be discussed in Sec.~\ref{subsubsec:422}.
\par
Again, Fig.~\ref{fig:1a} shows the PCs of the audio sound at $1$~kHz generated by circular RPs, evaluated for $20$ different combinations of design parameters, including the carrier frequencies, $f_{\mathrm{u,2}}$, and the ultra CD, $\mathscr{D}_{\mathrm{u,c}}$. The marked points indicate the critical audio SPL, $L_{p_{\mathrm{a,c}}}$, with different marker styles distinguishing the applied design parameters. At each $\mathscr{D}_{\mathrm{a,c}}$, a decrease in $f_{\mathrm{u,2}}$ or an increase in $\mathscr{D}_{\mathrm{u,c}}$ results in a higher $L_{p_{\mathrm{a,c}}}$. This relationship enables the determination of the optimal combination of $f_{\mathrm{u,0}}$ and $\mathscr{D}_{\mathrm{u,c}}$ that maximizes $L_{p_{\mathrm{a,c}}}$ at a given receiver location--that is, at a specific $\mathscr{D}_{\mathrm{a,c}}$. In this analysis, the surface normal velocity of the circular RP is fixed, as this represents an idealized scenario. However, this assumption does not hold in practical situations, where the transducer characteristics must be considered in detail, as further discussed in Sec.~\ref{subsec:23}.
\par
Figure~\ref{fig:1b} outlines the results of Fig.~\ref{fig:1a} in that the solid contour lines mean $L_{p_\mathrm{a,c}}$; the dashed contour lines indicate $\mathscr{D}_{\mathrm{a,c}}$; the dash-dotted lines are $a$. For the same audio CD, as the carrier frequency decreases, the audio SPL increases. For example, along the $\mathscr{D}_{\mathrm{a,c}} = 0.45$~m dashed contour line, as $f_{\mathrm{u,2}}$ decreases from $90$ to $40$~kHz, $L_{p_\mathrm{a,c}}$ increases approximately from $48$ to $53$~dB. For the same aperture radius, the audio SPL remains similar, but the audio CD varies with the carrier frequency. In the case of $a=50$ mm, increasing $f_{\mathrm{u,2}}$ from $40$ to $60$~kHz results in an increase in $\mathscr{D}_{\mathrm{a,c}}$ from $0.4$ to $0.5$~m. These relationships are further supported by the correlation matrix in Fig.~\ref{fig:s6}\cite{SM}.
\par
While prior studies have analyzed how aperture size and carrier frequency affect audio power in the far field,\cite{Zhong2021,Zhong2023} the present work instead focuses on the audio SPL at the audio CD, providing a practical guideline for selecting design parameters to match a target receiver location.

\subsection{\label{subsec:22} Acoustic characteristics of stepped plate}

\subsubsection{\label{subsubsec:221} Principle of stepped plate and its impedance characteristics}

\begin{figure}[ht]
    \begin{subfigure}[b]{8.5cm}
        \centering
        \includegraphics[width=\linewidth]{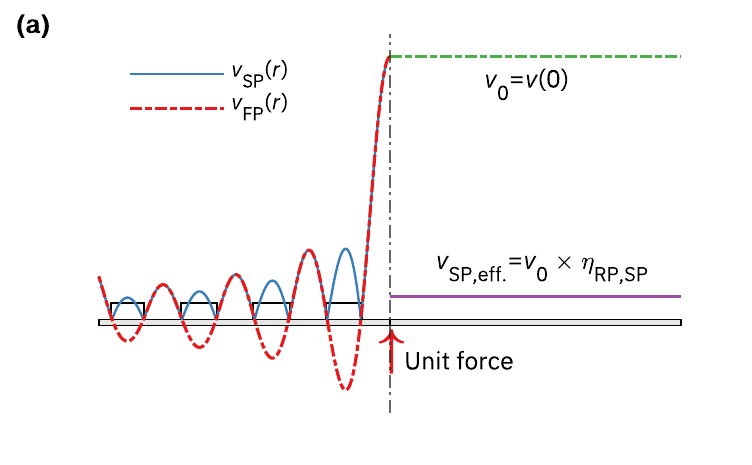}
        \caption{}
        \label{fig:2a}
    \end{subfigure}
    \begin{subfigure}[b]{8.5cm}
        \centering
        \includegraphics[width=\linewidth]{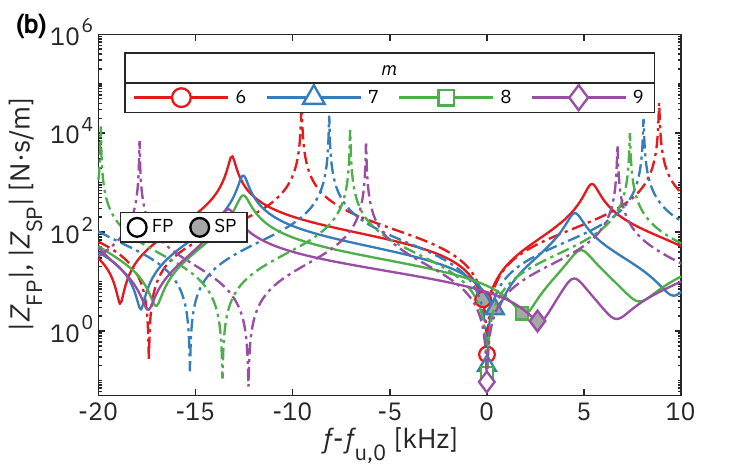}
        \caption{}
        \label{fig:2b}
    \end{subfigure}
    \caption{
        \label{fig:2}
        (a) Schematic comparison of a circular SP (left) and a RP (right). (b) Mechanical impedance at the plate center for FP (dash-dotted) and SP (solid), $Z_\mathrm{FP}$ and $Z_\mathrm{SP}$, respectively, from the 6th to 9th modes.
    }
\end{figure}

The SP, which is a circular FP modified with annular ring steps of half-wavelength height, $\lambda_\mathrm{u,0}/2$, is designed to approximate the coherent acoustic field of a uniformly vibrating RP.\cite{Barone1972} As shown in Fig.~\ref{fig:2a}, the dash-dotted line represents the SP's vibration mode shape, while the solid line approximates the surface particle velocity when phase-compensating steps are applied. If the center velocity of the SP is $v_0$, and its radiated field matches that of an RP, the equivalent RP surface velocity is given by $v_\mathrm{SP,eff} = v_0 \times \eta_{\mathrm{RP/SP}}$, where $\eta_{\mathrm{RP/SP}}$ denotes the ER. This section investigates the conditions under which the SP acoustically resembles the RP in axial propagation.
\par
Figure~\ref{fig:2b} presents the mechanical impedance at the center of the FP ($Z_\mathrm{FP}$) and SP ($Z_\mathrm{SP}$) from the 6th to 9th vibration modes, assuming a fixed FP resonance at $60$~kHz and an ultra CD of $0.45$~m, including radiation impedance in air. While the FP's resonance remains unchanged across modes, the addition of annular steps in the SP shifts the resonance--an effect more pronounced in higher-order modes--resulting in $Z_{\mathrm{SP}}$. In thin plates operating at higher-order modes, resonance shifts persist even after design refinement of the annular steps, suggesting that the added steps inevitably affect the resonance characteristics,\cite{Hwang2018} despite the use of polymer materials with relatively low density and stiffness.\cite{Kim2022} By contrast, thicker plates associated with lower-order modes exhibit reduced sensitivity to such modifications. Although not shown in full, similar trends were observed across all $120$ combinations of design parameters ($\mathscr{D}_\mathrm{u,c} = \{0.30, 0.35, 0.40, 0.45\}$~m, $f_{\mathrm{u,0}} = \{40, 50, 60, 75, 90\}$~kHz, and $m = \{4, 5, 6, 7, 8, 9\}$), simulated using COMSOL Multiphysics via LiveLink for MATLAB with frequency-domain acoustic-structure interaction analysis. For each case, the initial plate geometry was determined based on the target vibration mode using MMPT.\cite{Oh2023}

\subsubsection{\label{subsubsec:222} Evaluation of equivalence ratio at the stepped plate's resonant frequency}

\begin{figure}[ht]
    \begin{subfigure}[b]{8.5cm}
        \centering
        \includegraphics[width=\linewidth]{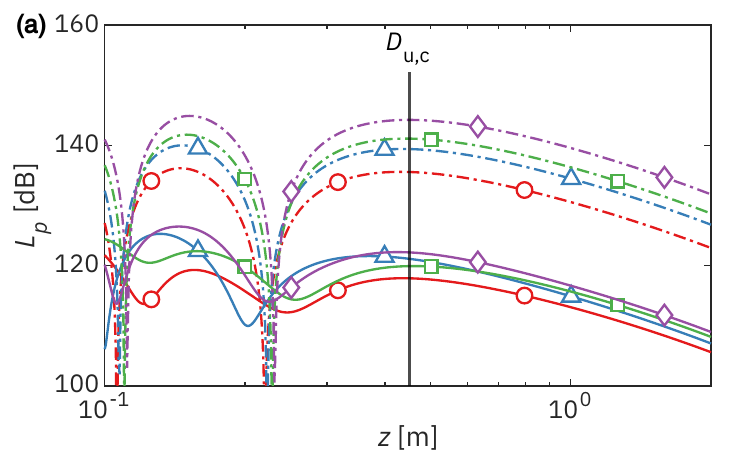}
        \caption{}
        \label{fig:3a}
    \end{subfigure}
    \begin{subfigure}[b]{8.5cm}
        \centering
        \includegraphics[width=\linewidth]{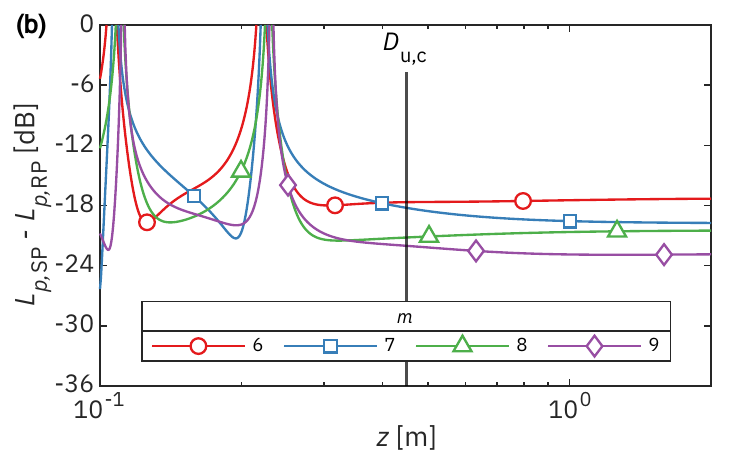}
        \caption{}
        \label{fig:3b}
    \end{subfigure}
    \caption{
        \label{fig:3}
        (a) Comparison of the PCs of acoustic fields generated by the SP (solid) and the RP (dash-dotted), and (b) its difference at the resonant frequency of the SP $f_{\mathrm{0,SP}}$.
    }
\end{figure}

As a next step, the PCs generated by the SP and the RP are compared. In the SP case, a unit force is applied at the center to excite the desired mode, and the resulting center velocity, $v_{\mathrm{0,SP}}$, is determined by the total mechanical impedance, including both structural and radiation components. For a fair comparison, the RP is assigned the same radius, and its surface velocity is set to match the center velocity of the SP, such that $v_{\mathrm{RP}} = v_{\mathrm{0,SP}}$.
\par
Figure~\ref{fig:3a} presents the PCs of the acoustic fields produced by the SP and RP at the SP's resonant frequency, $f_{\mathrm{0,SP}}$, which reflects the frequency shift introduced by the annular steps. Although the SP exhibits irregular behavior in the near field, both SP and RP PCs converge beyond $\mathscr{D}_\mathrm{u,c}$. The SPL difference at this distance defines the ER, $\eta_{\mathrm{SP/RP}}$, which quantifies the relative radiation efficiency. As shown in Fig.~\ref{fig:3b}, the ER depends on the ultra CD, $\mathscr{D}_\mathrm{u,c}$, the nominal ultrasound frequency $f_\mathrm{u,0}$, and the vibration mode $m$.
\par
For instance, the ER at the 8th mode is approximately $-20$~dB, implying that the RP must operate at a center velocity ten times lower than the SP to achieve similar far-field performance. Notably, even-mode SPs show consistent ER behavior along the acoustic axis beyond $\mathscr{D}_\mathrm{u,c}$, whereas odd-mode SPs exhibit fluctuations. This inconsistency is attributed to the absence of the outermost step in odd-mode designs due to fabrication difficulties. Moreover, the central region of the SP is left bare to avoid thermal degradation of the polymer steps during continuous operation.

\subsubsection{\label{subsubsec:223} Bandwidth characteristics of stepped plate}

\begin{figure}[ht]
    \begin{subfigure}[b]{8.5cm}
        \centering
        \includegraphics[width=\linewidth]{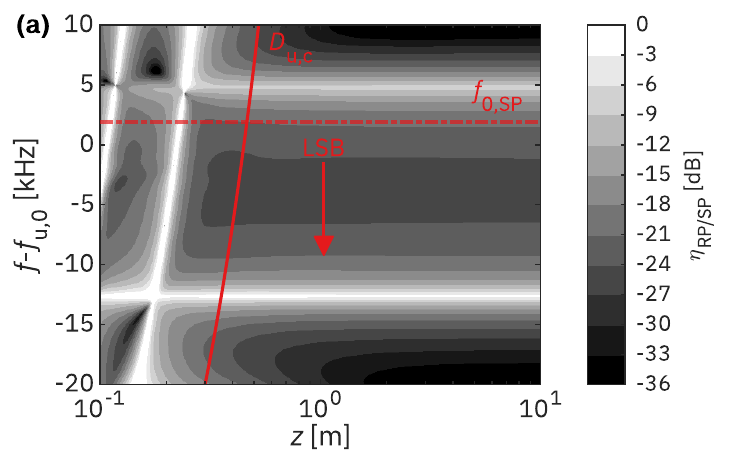}
        \caption{}
        \label{fig:4a}
    \end{subfigure}
    \begin{subfigure}[b]{8.5cm}
        \centering
        \includegraphics[width=\linewidth]{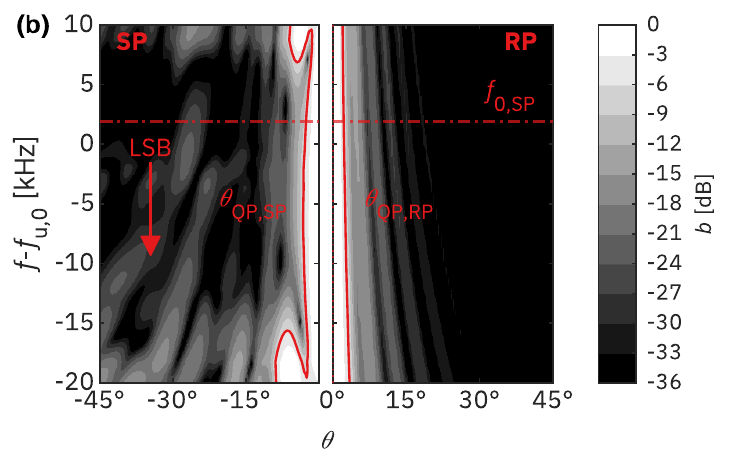}
        \caption{}
        \label{fig:4b}
    \end{subfigure}
    \caption{
        \label{fig:4}
        (a) Frequency-dependent variation of the ER $\eta_{\mathrm{SP/RP}}$ near the resonant frequency $f_\mathrm{u,0} = 60$~kHz and $\mathscr{D}_\mathrm{u,c} = 0.45$~m for the 8th mode. (b) BP comparison between the SP and RP measured at a distance of $1$~m. The left and right sides of the dotted vertical line correspond to the SP and RP, respectively.
    }
\end{figure}

For PAL applications, it is essential to evaluate whether the acoustic behavior of the SP aligns with that of a RP not only at the SP's resonant frequency, $f_{\mathrm{0,SP}}$, but also over a practical frequency bandwidth. This ensures consistent directional performance across the audio band. The present analysis focuses on the 8th vibration mode, with $f_\mathrm{u,0} = 60$~kHz and $\mathscr{D}_\mathrm{u,c} = 0.45$~m. Although omitted here for brevity, similar trends were observed across other combinations of design parameters, particularly for even-order modes (for addtional ressults, Figs.~\ref{fig:ERbandwidthOfSP} and \ref{fig:BPbandwidthOfSP}\cite{SM}).
\par
With LSB-AM modulation, the relevant frequency bandwidth extends approximately $10$~kHz below $f_{\mathrm{0,SP}}$. As shown in Fig.~\ref{fig:4a}, the ER varies within $3$~dB across this range, indicating robust axial field consistency. Additionally, although the BP of an SP at its design frequency is known to closely mimic that of an RP,\cite{Barone1972} its frequency-dependent behavior has not been thoroughly examined. Figure~\ref{fig:4b} presents BPs at $1$~m, with the left side representing the SP and the right side the RP. The SP maintains high directivity with a quarter-power angle, $\theta_{\mathrm{QP,SP}}$, of approximately $6$\textdegree\ throughout the bandwidth. Sidelobe levels remain below $-20$~dB, but begin to rise outside the range, approaching the main lobe level. This degradation is attributed to two main factors: frequency-dependent shifts in nodal circle locations that misalign the annular steps, and phase mismatch resulting from step heights optimized specifically for $f_{\mathrm{u,0}}$.
\par
These findings confirm that the ultrasonic acoustic fields generated by the RP and SP are comparable when both share the same radius and operating frequency. Although their surface velocity profiles differ, the similarity in PCs beyond $\mathscr{D}_\mathrm{u,c}$ and BPs suggests that the phase-compensating annular steps of the SP function effectively. Therefore, the preliminary design of an SP radiator can be guided using the well-established design principles of the RP.

\subsection{\label{subsec:23} Transducer characteristics of stepped plate for parametric array loudspeakers}

\begin{figure}[ht]
    \begin{subfigure}[b]{8.5cm}
        \centering
        \includegraphics[width=\linewidth]{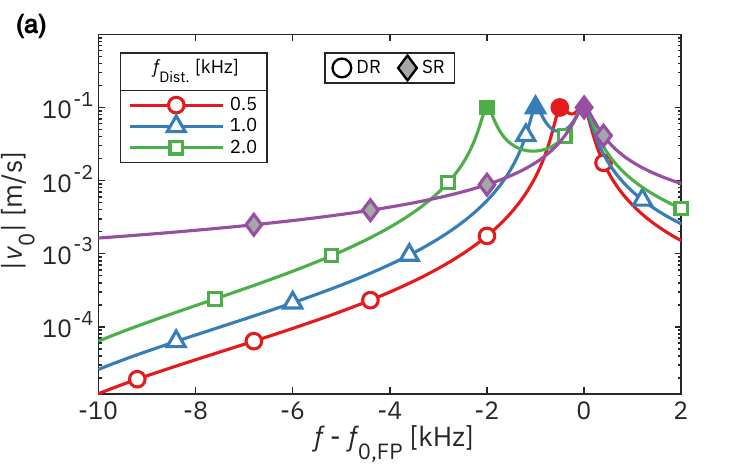}
        \caption{}
        \label{fig:5a}
    \end{subfigure}
    \begin{subfigure}[b]{8.5cm}
        \centering
        \includegraphics[width=\linewidth]{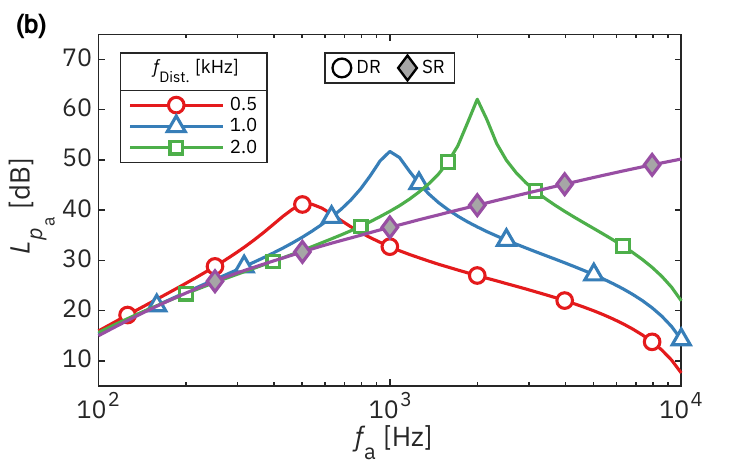}
        \caption{}
        \label{fig:5b}
    \end{subfigure}
    \caption{
        \label{fig:5}
        (a) Velocity frequency responses of hypothetical SR and DR ultrasonic transducers modeled in pole-zero-gain form. (b) Resulting audio SPL frequency responses at the audio CD $\mathscr{D}_{\mathrm{a,c}}$.
    }
\end{figure}

For PAL applications requiring a flat audio SPL response, an ideal approach involves employing an ultrasonic transducer with a flat frequency response and a $-12$~dB/oct. equalizer, as suggested by Berktay's far-field solution.\cite{Berktay1965} In practice, however, such transducers are difficult to realize, as most exhibit single resonance (SR) and lack broadband response.\cite{Zhong2024} The steep roll-off near resonance often exceeds $\pm 12$~dB/oct., causing poor low-frequency reproduction.\cite{Li2023, Zhong2022} To overcome this limitation, the SPPAL adopts a dual-resonance (DR) design that reinforces both carrier and sideband frequencies.\cite{Kim2022, Oh2023} The hypothetical velocity frequency responses of SR and DR transducers are modeled using a pole-zero-gain formulation:
\begin{align}
    v_{\mathrm{RP,SR}} &=  i\omega K \frac{1}{\omega_{\mathrm{r2}}^2(1+i\eta)-\omega^2} \\ 
    v_{\mathrm{RP,DR}} &=  i\omega K \frac{\omega_{\mathrm{a}}^2(1+i\eta)-\omega^2}{(\omega_{\mathrm{r1}}^2(1+i\eta)-\omega^2)(\omega_{\mathrm{r2}}^2(1+i\eta)-\omega^2)}
\end{align}

\noindent where $\omega_{\mathrm{r2}}$ and $\omega_{\mathrm{r1}}$ denote the angular frequencies of the carrier and sideband resonances, respectively; $\omega_\mathrm{a}$ is the anti-resonance angular frequency; and $\eta$ is the damping loss factor. These parameters are adjusted to reflect typical SPPAL behavior, as illustrated in Fig.~\ref{fig:5a}.
\par
As shown in Fig.~\ref{fig:5b}, compared to the SR transducer, the DR configuration exhibits noticeable amplification in the low audio band. The SPL curves, calculated using the SWE method for a RP with $f_\mathrm{u,0} = 60$~kHz and $\mathscr{D}_\mathrm{u,c} = 0.45$~m, confirm that DR design can compensate for low-frequency roll-off. Prior studies achieved a resonance spacing of about $3.3$~kHz,\cite{Kim2022, Oh2023} but achieving narrower spacing remains a challenge in practical implementation.

\section{\label{sec:3} Design}

In the previous section, the foundational design principles for PALs based on RPs were examined, followed by an exploration of the acoustic radiation characteristics of SPs, which approximate the behavior of RPs under certain conditions. This analysis highlighted the importance of carrier frequency, aperture size, and resonance mode selection in achieving optimal audio performance at a designated receiver location. Furthermore, the necessity of introducing DR in the ultrasonic velocity response of the transducer was emphasized to enhance low-frequency audio SPL. This section presents the design process of the SPPAL, where the SP is coupled with a Langevin-type transducer. Key design parameters are explored through multi-objective optimization, and their influence on audio performance is analyzed. Efficient simulation frameworks for nonlinear acoustic fields are introduced, followed by design-for-manufacturing considerations for practical implementation.

\subsection{\label{subsec:31} Prelminary design}

\subsubsection{\label{subsubsec:311} Design parameters and objective function behavior}

\begin{figure}[ht]
    \begin{subfigure}[b]{4.25cm}
        \centering
        \includegraphics[width=\linewidth]{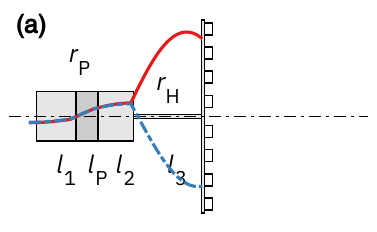}
        \caption{}
        \label{fig:6a}
    \end{subfigure}
    \begin{subfigure}[b]{4.25cm}
        \centering
        \includegraphics[width=\linewidth]{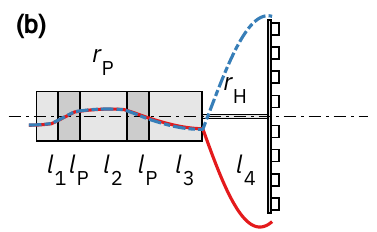}
        \caption{}
        \label{fig:6b}
    \end{subfigure}
    \\
    \begin{subfigure}[b]{4.25cm}
        \centering
        \includegraphics[width=\linewidth]{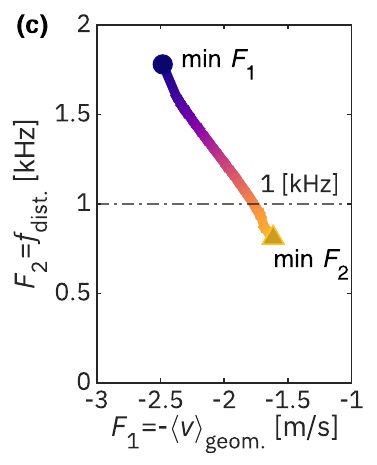}
        \caption{}
        \label{fig:6c}
    \end{subfigure}
    \begin{subfigure}[b]{4.25cm}
        \centering
        \includegraphics[width=\linewidth]{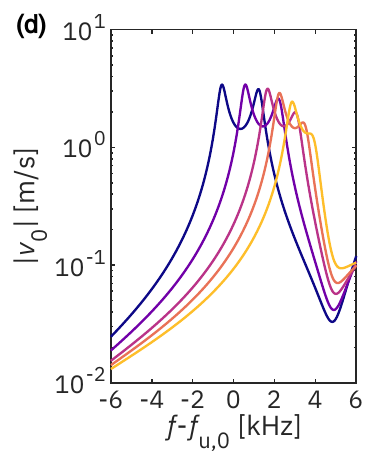}
        \caption{}
        \label{fig:6d}
    \end{subfigure}
    \caption{
        \label{fig:6}
        Schematic of (a) a half-wavelength Langevin transducer with a single piezoelectric stack, and (b) a full-wavelength transducer with a cascaded double-stack configuration, showing the fundamental mode ($f_\mathrm{r1}$, red solid line) and the secondary mode ($f_\mathrm{r2}$, blue dash-dotted line), respectively. (c) Pareto front showing the relationship between objective functions $F_1$ and $F_2$, illustrating the trade-off in DR optimization. (d) Frequency responses of plate center velocity obtained from multi-objective optimization. The darkest and lightest lines represent the optimized responses corresponding to $\min F_1$ and $\min F_2$, respectively. 
    }
\end{figure}

When the frequency distance between the fundamental mode and its secondary mode of the horn-coupled Langevin transducer is small and those resonant frequencies are also close to the intended mode's resonant frequency of the stepped plate, DR occurs closely. Thus, the key to causing DR is the modal transition that occurs with small frequency variations. To realize this effect in the SPPAL design, various design parameters, denoted by $\mathbf{p}$, were considered:

\begin{itemize}
    \item 
    Transducer configuration:\\
    $\mathscr{C}_\mathrm{Xdcr}=\{\text{\textquotedblleft Half\textquotedblright},\text{\textquotedblleft Full\textquotedblright}\}$
    \item
    Radial dimension of the transducer's piezo stack:\\
    $r_\mathrm{P}=\{x|\lambda_\mathrm{P,1}/8<x<\lambda_\mathrm{P,1}/4\}$ where $\lambda_\mathrm{P,1}=v_1^E/f_\mathrm{u,0}$
    \item
    Length of the transducer's piezo stack:\\
    $l_\mathrm{P}=\{x|\lambda_\mathrm{P,3}/10<x<\lambda_\mathrm{P,3}/4\}$ where $\lambda_\mathrm{P,3}=v_3^E/f_\mathrm{u,0}$
    \item
    Horn-end radius of the stepped horn:\\
    $r_\mathrm{H}=\{0.75,1.00,1.25,1.50\}$~mm
\end{itemize}

\noindent Here, $\mathscr{C}_\mathrm{Xdcr}$ specifies the transducer's fundamental mode configuration. The half-wavelength transducer uses a single piezoelectric stack (Fig.~\ref{fig:6a}), whereas the full-wavelength configuration consists of a cascaded double-stack (Fig.~\ref{fig:6b}). In both schematics, the red solid line indicates the fundamental mode ($f_\mathrm{r1}$), and the blue dash-dotted line represents the secondary mode ($f_\mathrm{r2}$). The parameters $r_\mathrm{P}$ and $l_\mathrm{P}$ are normalized to the material-dependent wavelengths in the radial and longitudinal directions, respectively, with wave speeds defined as $v_i^E = (1/\rho s_{ii}^E)^{1/2}$, where $\rho$ is the material density and $s_{ii}^E$ is the elastic compliance under a constant electric field for direction $i$ (1: radial, 3: longitudinal). The horn-end radius $r_\mathrm{H}$ is constrained by manufacturability--smaller values yield thin-neck geometries that are challenging to fabricate.
\par
At the preliminary design stage, a forward design approach is adopted to assess how transducer design parameters and configurations affect key characteristics such as the frequency spacing of DR ($f_\mathrm{dist.}$) and the corresponding modal velocities ($v_\mathrm{r1}$ and $v_\mathrm{r2}$), using multi-objective optimization. Since this process is computationally intensive, the objective function must be both efficient to evaluate and physically meaningful.
\par
For transducer sections with diameters smaller than a quarter wavelength, longitudinal motion dominates, and the 1D Langevin model provides sufficient accuracy.\cite{Neppiras1973,RanzGuerra1975} However, to achieve tighter frequency spacing (e.g., $f_\mathrm{dist.} \sim 1$~kHz), larger-diameter sections are often necessary, where the 1D model becomes inadequate due to radial-longitudinal coupling. To address this, an approximate three-dimensional (Approx. 3D) model incorporating Poisson effects was employed, enabling accurate mode prediction without resorting to full FEM modeling.\cite{Iula1998}
\par
In the multi-objective optimization process, the design parameters, $\mathbf{p}$, are listed above and the design variables, $\mathbf{x}$, to be optimized are the length of each part of the transducers, $\mathbf{x} = \{l_1, l_2, l_3 | \mathscr{C}_\mathrm{Xdcr} = \text{\textquotedblleft Half\textquotedblright}\}; \mathbf{x} = \{l_1, l_2, l_3, l_4 | \mathscr{C}_\mathrm{Xdcr} = \text{\textquotedblleft Full\textquotedblright}\}$, its lower and upper bounds are set to $0.5$ to $1.5$ times those of Langevin's equation,\cite{Neppiras1973,RanzGuerra1975} also two objective functions are defined:

\begin{itemize}
    \item 
    Objective function 1,\\
    $F_1=-\langle v\rangle_\mathrm{geom.}=-(v_\mathrm{r1}\cdot v_\mathrm{r2}\cdot v_\mathrm{m} )^{1/3}$, is the negative geometric mean of the two highest peaks, $v_\mathrm{r1}$ and $v_\mathrm{r2}$, and the local minimum $v_\mathrm{m}$ between them in the plate's center velocity.
    \item
    Objective function 2,\\
    $F_2=f_\mathrm{dist.}=f_\mathrm{r2} - f_\mathrm{r1}$, is the frequency difference between the two largest peaks.
\end{itemize}

\noindent Multi-objective optimization was performed through the \texttt{gamultiobj} function, with \texttt{useParallel=true} set by Global Optimization Toolbox and Parallel Computing Toolbox in MATLAB 2024b. The specific procedure used for this process is summarized in Algorithm~\ref{alg:multiopt}\cite{SM}. The optimization produced a Pareto front, illustrating the trade-offs between the two objectives as shown in Fig.~\ref{fig:6c}. Figure~\ref{fig:6d} presents the frequency responses of the plate's center velocity when either $F_1$ or $F_2$ is minimized. When the minimization of $F_1$ is prioritized, the velocity amplitudes at both resonances become comparable ($v_\mathrm{r1} \approx v_\mathrm{r2}$), as depicted by the darkest line. In contrast, minimizing $F_2$ yields a smaller frequency difference, but the velocity amplitude at $f_\mathrm{r2}$ is generally lower than at $f_\mathrm{r1}$ ($v_\mathrm{r1} > v_\mathrm{r2}$), as indicated by the lightest line. While only five representative cases are shown, the frequency response gradually changes along the Pareto front as the optimization objective shifts from $\min F_1$ to $\min F_2$.
\par
In summary, when $F_2$ is minimized, $f_\mathrm{dist.}$ decreases, but $v_\mathrm{r2}$ decreases, so the audio SPL of the PAL decreases overall due to the low carrier velocity magnitude in the case of employing LSB-AM. In contrast, when is $F_1$ minimized, the velocity of the DR is maximized and $f_\mathrm{dist.}$ is minimized as possible for the corresponding design parameter. Therefore, considering the characteristics of the frequency response curve of SPPAL, $\min F_1$ is the most optimal choice. The design parameters corresponding to the results shown in Fig.~\ref{fig:6c}-\ref{fig:6d} are as follows: the stepped plate of $\mathscr{D}_\mathrm{u,c}=0.45$~m, $f_\mathrm{u,0}=60$~kHz, and 8th mode is coupled to the half-wavelength transducer of $r_\mathrm{P}=9$~mm, $l_\mathrm{P}=8$~mm, and $r_\mathrm{H}=0.75$ mm, and the results as not shown here follow similar trends for the other design parameters.

\subsubsection{\label{subsubsec:312} Feasibility of dual-resonance across design parameter combinations}

\begin{figure}[ht]
    \begin{subfigure}[b]{4.25cm}
        \centering
        \includegraphics[width=\linewidth]{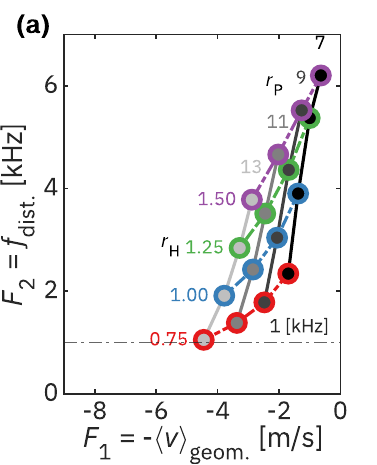}
        \caption{}
        \label{fig:7a}
    \end{subfigure}
    \begin{subfigure}[b]{4.25cm}
        \centering
        \includegraphics[width=\linewidth]{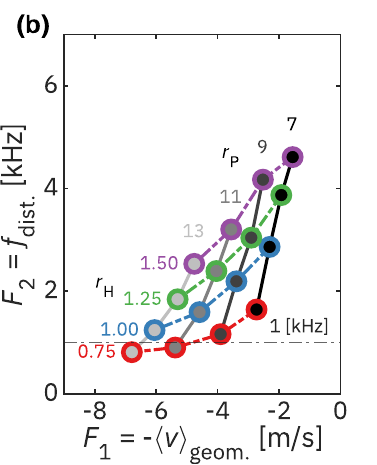}
        \caption{}
        \label{fig:7b}
    \end{subfigure}
    \caption{
        \label{fig:7}
        Multi-objective optimization results for (a) $\mathscr{C}_\mathrm{Xdcr} = \text{\textquotedblleft Half\textquotedblright}$ and (b) $\mathscr{C}_\mathrm{Xdcr} = \text{\textquotedblleft Full\textquotedblright}$, showing the relationship between $F_1$ and $F_2$ across different design parameters.
    }
\end{figure}

The design parameter range for the transducer coupled to the stepped plate considered above is defined as $\mathscr{C}_\mathrm{Xdcr} = \{ \text{\textquotedblleft Half\textquotedblright}, \text{\textquotedblleft Full\textquotedblright} \}$, $r_\mathrm{P} = \{7, 9, 11, 13\}$~mm, $r_\mathrm{H} = \{0.75, 1.00, 1.25, 1.50\}$~mm, and $l_\mathrm{P} = 8$~mm. The corresponding multi-objective optimization results are summarized in Fig.~\ref{fig:7}. Solid lines connect points sharing the same $r_\mathrm{P}$, while dash-dotted lines connect those with the same $r_\mathrm{H}$. As $r_\mathrm{P}$ and $r_\mathrm{H}$ vary, both $F_1$ and $F_2$ show monotonic trends, suggesting that interpolation may be used to estimate intermediate values despite the discrete sampling of parameters.
\par
Certain combinations of parameters satisfy the target condition $f_\mathrm{dist.}=1$~kHz, while others do not. For example, in the half-wavelength configuration, this condition is only satisfied when $r_\mathrm{H} = 0.75$~mm and $r_\mathrm{P} = 13$~mm. In comparison, the full-wavelength configuration achieves similar results for $r_\mathrm{H} = 0.75$~mm with $r_\mathrm{P} = \{9, 11\}$~mm, or $r_\mathrm{H} = 1.00$~mm with $r_\mathrm{P} = 13$~mm. However, a small horn-end radius $r_\mathrm{H}$--especially in thin-necked geometries as illustrated in Figs.~\ref{fig:6a} and \ref{fig:6b}--can present fabrication challenges due to potential structural weakness. Therefore, manufacturability must be considered in selecting feasible design combinations.
\par
In summary, DR operation with $f_\mathrm{dist.}=1$~kHz is more easily achieved with the full-wavelength configuration than with the half-wavelength configuration. This condition tends to be met when $r_\mathrm{H}$ is small and $r_\mathrm{P}$ is large.

\subsection{\label{subsec:32} Capability of SPPAL for audio sound pressure level}

\begin{figure*}[ht]
    \begin{subfigure}[b]{8.5cm}
        \centering
        \includegraphics[width=\linewidth]{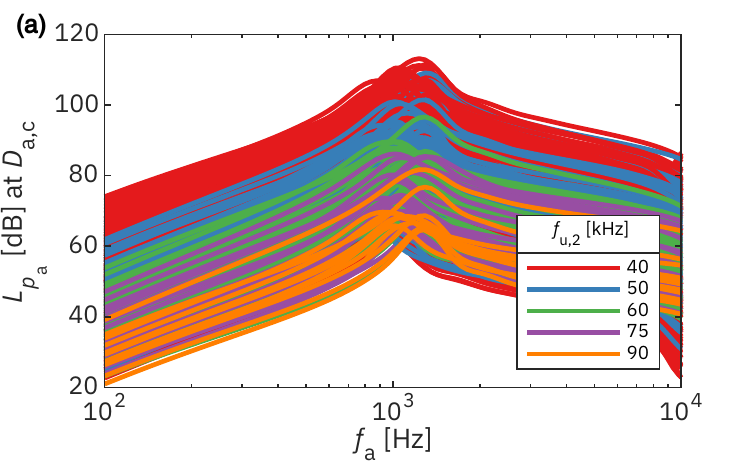}
        \caption{}
        \label{fig:8a}
    \end{subfigure}
    \begin{subfigure}[b]{8.5cm}
        \centering
        \includegraphics[width=\linewidth]{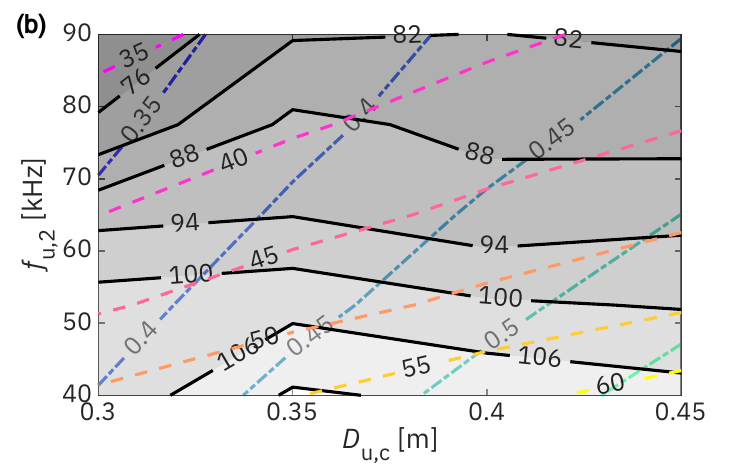}
        \caption{}
        \label{fig:8b}
    \end{subfigure}
    \\
    \begin{subfigure}[b]{17.8cm}
        \centering
        \includegraphics[width=\linewidth]{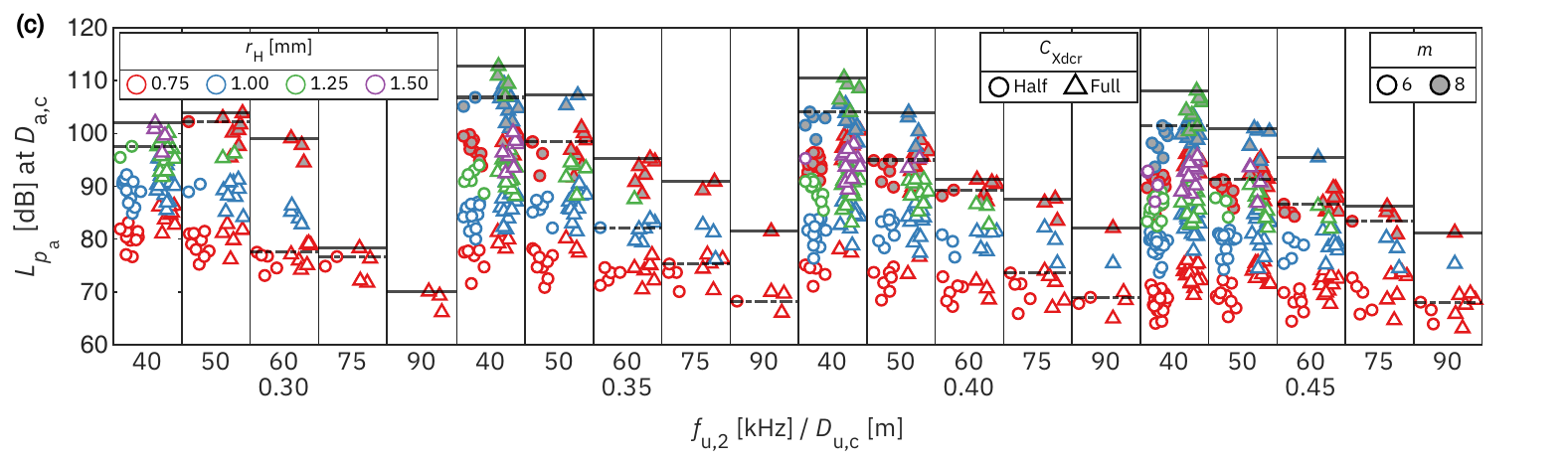}
        \caption{}
        \label{fig:8c}
    \end{subfigure}
    \caption{
        \label{fig:8}
        Nonlinear acoustic analysis results for the SPPAL obtained using the SWE method (Method A). (a) Frequency response of the audio CD $\mathscr{D}_{\mathrm{a,c}}$. (b) Summary contour plot showing the design parameter: solid lines indicate the critical audio SPL $L_{p_\mathrm{a,c}}$, dashed lines indicate the $\mathscr{D}_{\mathrm{a,c}}$, and dash-dotted lines denote the aperture radius $a$. (c) Peak audio SPL at $\mathscr{D}_{\mathrm{a,c}}$ for various design parameter combinations of the SPPAL.
    }
\end{figure*}

The center velocity frequency response of the plate, $v_0$, in Fig.~\ref{fig:6d}, is converted to the effective surface normal velocity of the stepped plate, $v_\mathrm{SP,eff.}$, using the ER, $\eta_\mathrm{SP/RP}$. Among the DR SPPALs designed via multi-objective optimization, the audio SPL at the audio CD $L_{p_\mathrm{a}}(\mathscr{D}_\mathrm{a,c})$ is predicted using the SWE method, as further detailed in Sec.~\ref{subsec:s21}\cite{SM}. The simulation time per case is approximately $20$ seconds, enabling rapid exploration across a broad parameter space. Figure~\ref{fig:8a} presents the results for cases with $f_\mathrm{dist.} = \{x \mid 800 < x < 1250\}$~Hz.
\par
Figure~\ref{fig:8c} summarizes these results by presenting the peak audio SPL at audio CD as a representative performance metric, offering a compact visualization of how the design parameters influence acoustic output. Each box corresponds to a specific combination of $\mathscr{D}_\mathrm{u,c}$ and $f_\mathrm{u,2}$. White and gray-faced markers indicate the 6th and 8th modes, respectively, while circular and triangular markers represent half- and full-wavelength transducer. Within each box, solid and dash-dotted horizontal lines indicate the maximum audio SPL for the full- and half-wavelength configurations, respectively. The color of each marker encodes the horn-end radius, $r_\mathrm{H}$. This summary plot facilitates an intuitive understanding of the performance landscape across the design space and provides practical guidance for SPPAL design. Although certain design parameters were expected to involve trade-offs, the evaluated parameter space shows generally monotonic trends without clear evidence of internal optima.
\par
Several such trade-offs can be anticipated. For example, with a fixed $\mathscr{D}_\mathrm{u,c}$, increasing $f_\mathrm{u,0}$ reduces the aperture radius $a$, which decreases the mechanical impedance $Z_\mathrm{SP}$ and increases $v_\mathrm{SP,eff.}$--enhancing primary wave radiation. However, the resulting reductions in both aperture size and absorption length lead to a smaller effective volume of the parametric array, thereby limiting the generation of difference-frequency components and potentially lowering the audio SPL. Similarly, increasing the plate's mode number reduces its thickness, which lowers $Z_\mathrm{SP}$ and may improve $v_\mathrm{SP,eff.}$, but at the expense of a reduced ER--again potentially diminishing the overall audio SPL.
\par
Furthermore, Fig.~\ref{fig:8c} allows estimation of the maximum horn-end radius $r_\mathrm{H}$ that can satisfy the $f_\mathrm{dist.} \approx 1$~kHz condition for various design parameter combinations. The results suggest that full-wavelength configurations can generally accommodate slightly larger $r_\mathrm{H}$ values than half-wavelength ones, improving manufacturability without compromising acoustic performance. As $r_\mathrm{H}$ directly affects the ease and cost of fabrication--particularly in avoiding thin-necked horns--identifying the upper bound of $r_\mathrm{H}$ is critical for cost-effective implementation.
\par
It is noted that some simulation results may exhibit reduced accuracy due to the limitations of the Westervelt-based SWE model, which excludes effects such as Lagrangian density considered in the Kuznetsov equation. However, these findings should thus be interpreted as indicative trends rather than precise quantitative predictions.
\par
Finally, Fig.~\ref{fig:8b} visualizes a contour map summarizing the results in Fig.~\ref{fig:8c}, overlaid on the parameter space of $\mathscr{D}_\mathrm{u,c}$ and $f_\mathrm{u,2}$. This representation corresponds to the SPPAL version shown in Fig.~\ref{fig:1b}, offering a compact overview of design trends. Solid contour lines indicate $L_{p_\mathrm{a,c}}$, dash-dotted lines show constant $\mathscr{D}_\mathrm{a,c}$, and dashed lines denote the aperture radius $a$. For example, at $\mathscr{D}_\mathrm{a,c} = 0.45$~m, decreasing $f_\mathrm{u,2}$ from $90$ to $40$~kHz raises $L_{p_\mathrm{a}}$ from approximately $82$ to $106$~dB, guiding appropriate SPPAL configurations for target receiver locations. These interdependencies among design parameters and performance metrics can be further understood by referring to the correlation matrix in Fig.\ref{fig:s7}\cite{SM}.

\subsection{\label{subsec:33} Design for manufacturing}

The design guide developed in Sec.~\ref{subsec:32} serves as a basis for selecting appropriate parameters according to target audio performance and receiver placement. This section focuses on a design-for-manufacturing (DFM) approach, in which the selected parameters are refined with practical constraints in mind. In addition to performance considerations, DFM requires attention to structural integrity under assembly and operational loads, including factors such as thread stripping during piezoelectric prestressing and the linear elastic limit of the preload bolt.\cite{Erik2020, DeAngelis2015} Accordingly, key aspects such as thread engagement length, piezoelectric ring geometry, and fastener dimensions must be optimized.\cite{Neppiras1973}

\subsubsection{\label{subsubsec:331} Design through global optimization}

\begin{figure}[ht]
    \centering
    \includegraphics[width=\linewidth]{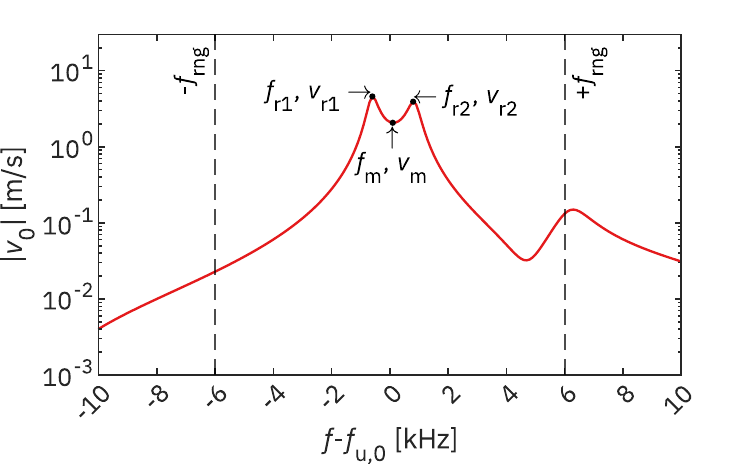}
    \caption{
        \label{fig:9}
        Center velocity frequency response of the optimized SPPAL under acoustic loading conditions. The result corresponds to the globally optimized design with $F_1 = -3.33$~m/s and $F_2 = 1.4$~kHz.
    }
\end{figure}

Based on the discussion in Sec.~\ref{subsubsec:311}, global optimization was conducted under the premise that minimizing only objective function~1, $F_1$, is sufficient to achieve the desired velocity frequency response of the SPPAL. Although surrogate optimization algorithms do not require the objective function to be smooth, they perform optimally when the function is continuous.\cite{MathWorks2024} Additionally, because reducing computational cost is prioritized in FEM-based optimization, objective function evaluations were conducted according to Algorithm~\ref{alg:globalopt}\cite{SM}, which introduces intentional discontinuities without significantly impairing optimization performance. Additionally, acoustic loading was excluded during optimization to further reduce computational effort. Instead, acoustic simulations were conducted afterward to evaluate the optimized design.
\par
The final design parameters selected from global optimization are as follows: $\mathscr{D}_\mathrm{u,c} = 0.45$~m, $f_\mathrm{u,0} = 60$~kHz, $m = 8$, $\mathscr{C}_\mathrm{Xdcr} = \text{\textquotedblleft Full\textquotedblright}$, $r_\mathrm{P} = 9$~mm, $l_\mathrm{P} = 8$~mm, and $r_\mathrm{H} = 0.75$~mm. The piezoelectric ring was fabricated using C-21 material (Fuji Ceramics), with an inner diameter $r_\mathrm{P,i} = 2.5$~mm, thickness $t_\mathrm{P} = 2$~mm, and a copper electrode thickness $t_\mathrm{elec} = 150$~\textmu m. An M4$\times$0.7 metric fastener was employed, with a counterbore depth $d_\mathrm{cb} = 0.5$~mm, complete thread length $l_\mathrm{ct} = 5$~mm, incomplete thread length $l_\mathrm{ict} = 3$~mm, and minimum remaining wall thickness $t_\mathrm{rem} = 3$~mm. The optimized SPPAL's center velocity frequency response under acoustic loading conditions is shown in Fig.~\ref{fig:9}, with $F_1 = -3.33$~m/s and $F_2 = 1.4$~kHz.
\par
Global optimization was performed using the \texttt{surrogateopt} function from the Global Optimization Toolbox in MATLAB 2024b. Design variables and simulation results were exchanged with COMSOL Multiphysics via LiveLink for MATLAB. Each objective function evaluation involved eigenfrequency and frequency-domain analyses incorporating piezoelectric-solid and electrostatic coupling. The entire process took approximately three hours on a workstation equipped with dual Intel Xeon Silver 4210 processors ($2.20$~GHz) and $384$~GB of RAM.

\subsubsection{\label{subsubsec:332} Framework for nonlinear acoustic simulation}

To further evaluate the nonlinear acoustic performance of the DFM-optimized SPPAL, two numerical frameworks were considered: a hybrid approach combining FEM with the SWE method (Method B), and a full-domain FEM simulation (Method C). Both are based on successive approximations of the Westervelt equation but differ significantly in computational cost and spatial coverage. Method B enables efficient evaluation of the audio beam pattern by using the FEM-computed velocity distribution of the stepped plate as the boundary condition for SWE, whereas Method C directly simulates the full nonlinear field, requiring high spatial resolution and large computational domains. A detailed comparison between these methods is provided in Sec.~\ref{sec:s2}\cite{SM}, demonstrating that Method B offers a favorable trade-off between accuracy and efficiency, particularly for modeling directional audio fields in resource-constrained environments.

\section{\label{sec:4} Experiment}

\subsection{\label{subsec:41} Realization}

\begin{figure}[ht]
    \begin{subfigure}[b]{4.25cm}
        \centering
        \includegraphics[width=\linewidth]{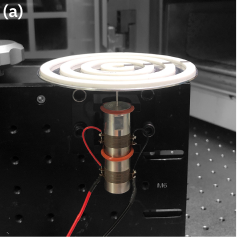}
        \caption{}
        \label{fig:10a}
    \end{subfigure}
    \begin{subfigure}[b]{4.25cm}
        \centering
        \includegraphics[width=\linewidth]{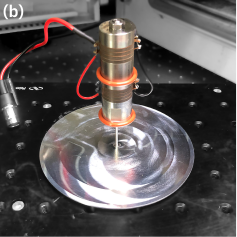}
        \caption{}
        \label{fig:10b}
    \end{subfigure}
    \caption{
        \label{fig:fabricated_xdcr}
        Photographs of the fabricated stepped plate parametric array loudspeaker: (a) top view showing the polymer annular step ring, and (b) bottom view showing the aluminum stepped plate and assembled transducer.
    }
\end{figure}

Figure~\ref{fig:fabricated_xdcr} shows photographs of the fabricated SPPAL from top and bottom views. The transducer components were machined from aluminum and stainless steel. During assembly, a custom-designed clamping jig was used to hold the transducer in place under bias, while a dedicated aligner ensured accurate coaxial alignment between the piezoelectric rings and the transducer body. The preload on the bolt was controlled using a Tohnichi CL15NX8D torque wrench to maintain consistent mechanical prestress. The polymer annular step ring was fabricated following the procedure detailed in previous studies.\cite{Kim2022, Oh2023} Specifically, the polymer step was cast using a silicone mold shaped as an annular ring. The mold was filled with a mixture of Henkel Loctite Stycast 1266 epoxy encapsulant and PB-725, a micro-shell-type ultralight filler supplied by Seokyung CMT, at a volume mixing ratio of 1:3.

\subsection{\label{subsec:42} Vibrational characteristics}

\begin{figure}[ht]
    \centering
    \includegraphics[width=\linewidth]{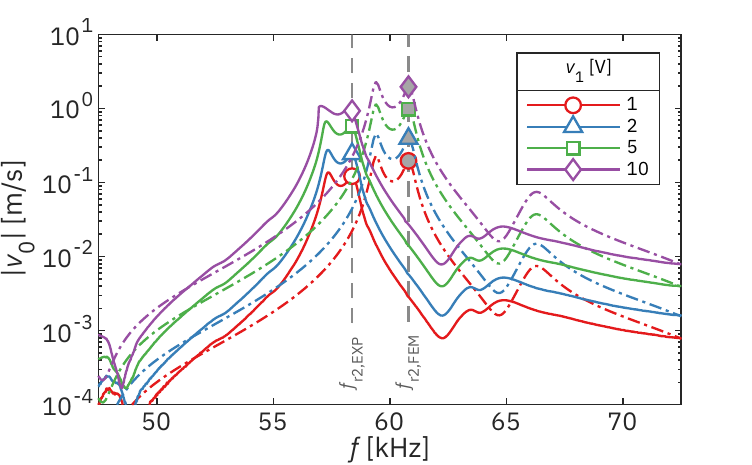}
    \caption{
        \label{fig:vib_fr}
        Measured and simulated velocity frequency responses at the center of the stepped plate under driving voltages of $1$, $2$, $5$, and $10$~V. Solid lines represent experimental measurements, while dash-dotted lines correspond to FEM simulations based on the DFM-optimized design.
    }
\end{figure}

\subsubsection{\label{subsubsec:421} Frequency response of plate's center velocity}

Figure~\ref{fig:vib_fr} presents the measured frequency response of the plate's center velocity under driving voltages of $1$, $2$, $5$, and $10$~V, spanning the $40-80$~kHz range. Solid lines represent experimental data, while dash-dotted lines correspond to FEM results based on the DFM-optimized design. The simulations assumed an applied voltage of $20$~V (equivalent to $10$~kV/m), whereas in practice, the transducer was driven up to $10$~V due to the onset of nonlinear effects.
\par
The measurements clearly demonstrate DR behavior near $60$~kHz, in qualitative agreement with FEM predictions. Quantitatively, $F_1$ reached approximately $2$~m/s at the maximum driving voltage of $10$~V, while $F_2$ was around $1$~kHz, with resonance peaks between $57.3$~kHz and $58.3$~kHz. Vertical dashed lines labeled $f_{\mathrm{r2,FEM}}$ and $f_{\mathrm{r2,EXP}}$ indicate the simulated and measured carrier frequencies for LSB-AM operation. A downward frequency shift was observed due to fabrication tolerances, which resulted in a thinner-than-designed plate. At voltages above $10$~V, nonlinear effects emerged, including amplitude growth and resonance shifts. These results highlight the need to consider both nonlinear dynamics and manufacturing accuracy in practical implementation. Further refinement of the piezoelectric ring and preload bolt design is recommended to reduce such discrepancies.

\subsubsection{\label{subsubsec:422} Experimental modal analysis}

\begin{figure}[ht]
    \begin{subfigure}[b]{4.25cm}
        \centering
        \includegraphics[width=\linewidth]{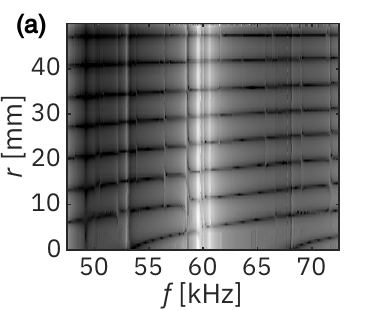}
        \caption{}
        \label{fig:vib_ma_fp_ampli}
    \end{subfigure}
    \begin{subfigure}[b]{4.25cm}
        \centering
        \includegraphics[width=\linewidth]{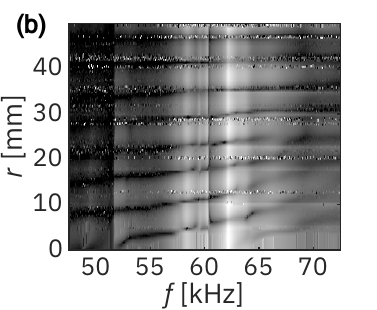}
        \caption{}
        \label{fig:vib_ma_sp_ampli}
    \end{subfigure}
    \\
    \begin{subfigure}[b]{4.25cm}
        \centering
        \includegraphics[width=\linewidth]{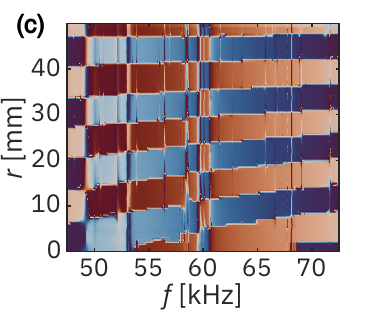}
        \caption{}
        \label{fig:vib_ma_fp_phase}
    \end{subfigure}
    \begin{subfigure}[b]{4.25cm}
        \centering
        \includegraphics[width=\linewidth]{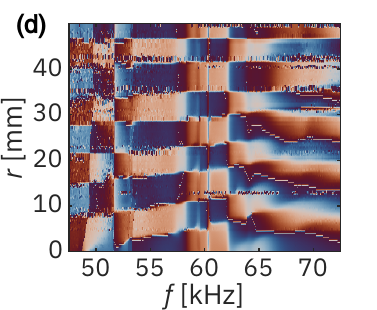}
        \caption{}
        \label{fig:vib_ma_sp_phase}
    \end{subfigure}
    \caption{
        \label{fig:vib}
        Experimental modal analysis results of the stepped plate transducer. (a), (c) Normalized amplitude and phase distributions along the radial direction for the FP without annular ring steps. (b), (d) Corresponding normalized amplitude and phase results for the stepped plate. In the phase plots, the colors are mapped such that white represents $0$, while dark blue and red correspond to $-\pi$ and $\pi$, respectively.
    }
\end{figure}

Figure~\ref{fig:vib} presents the experimental modal analysis (EMA) results of the SPPAL, showing amplitude and phase distributions along the radial direction. Figures~\ref{fig:vib_ma_fp_ampli} and \ref{fig:vib_ma_fp_phase} depict the response of a FP without annular steps, while Figs.~\ref{fig:vib_ma_sp_ampli} and \ref{fig:vib_ma_sp_phase} correspond to the SP. The FP vibrates clearly in the 8th flexural mode between approximately $53$ and $67$~kHz, with eight distinct phase reversals. Mode transitions to the 7th and 9th modes occur below and above this range, respectively. In contrast, the SP exhibits an earlier shift to the 9th mode near $63$~kHz, attributed to the mechanical impedance changes introduced by the annular steps. The frequency span of the 8th mode defines the effective operational bandwidth of the SPPAL, over which it approximates the coherent radiation of a RP. While the specimen used for EMA was a different sample from the one analyzed earlier, it followed the same design; thus, despite slight frequency shifts, the modal behavior and step-induced effects remain consistent.
\par
These modal characteristics play a crucial role in selecting the modulation algorithm for SPPAL operation. In particular, LSB-AM is identified as the optimal strategy. By selecting the higher resonance (approximately $61$~kHz) as the carrier, LSB-AM ensures a broader usable LSB, compared to the upper sideband (USB). In contrast, modulation schemes such as double-sideband amplitude modulation (DSB-AM) suffer from poor USB bandwidth. Wideband approaches like square-root and modified amplitude modulation (SRT-AM, MAM)\cite{Zhong2024} are also unsuitable due to the limited radiation bandwidth inherent to SPPAL.

\subsection{\label{subsec:43} Acoustic characteristics}

\subsubsection{\label{subsubsec:431} Ultrasound}

\paragraph{\label{paragraph:4311} Frequency response}

\begin{figure}[ht]
    \centering
    \includegraphics[width=\linewidth]{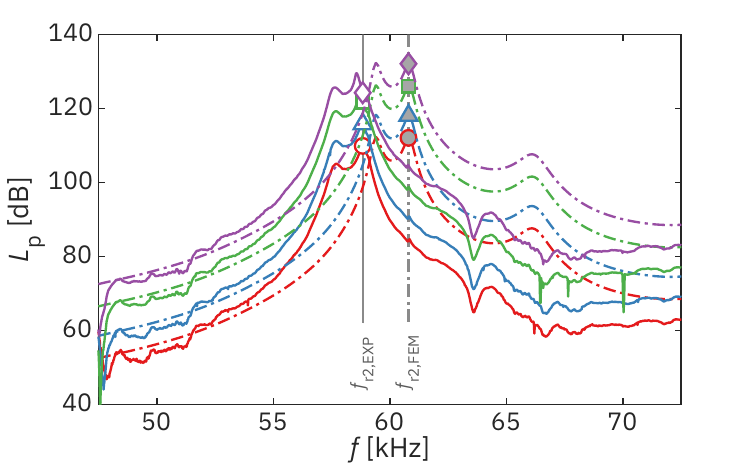}
    \caption{
        \label{fig:ultra_fr}
        Measured and simulated ultrasound frequency responses at the ultra CD ($\mathscr{D}_\mathrm{u,c} = 0.45$~m) under driving voltages of $1$, $2$, $5$, and $10$~V. Solid lines represent experimental measurements, while dash-dotted lines correspond to FEM simulations based on the DFM-optimized design.
    }
\end{figure}

Figure~\ref{fig:ultra_fr} presents the measured ultrasound frequency responses of SPPAL at driving voltages of $1$, $2$, $5$, and $10$~V, recorded at $\mathscr{D}_\mathrm{u,c} = 0.45$~m. Solid lines represent experimental data, while dash-dotted lines indicate FEM simulation results from Method B. Similar to Fig.~\ref{fig:vib_fr}, the ultrasound acoustic responses exhibit pronounced DR peaks around the experimentally determined carrier frequency $f_\mathrm{r2,EXP}\approx 60$ kHz.

\paragraph{\label{paragraph:4312} Propagation curve}

\begin{figure*}[ht]
    \begin{subfigure}[b]{8.5cm}
        \centering
        \includegraphics[width=\linewidth]{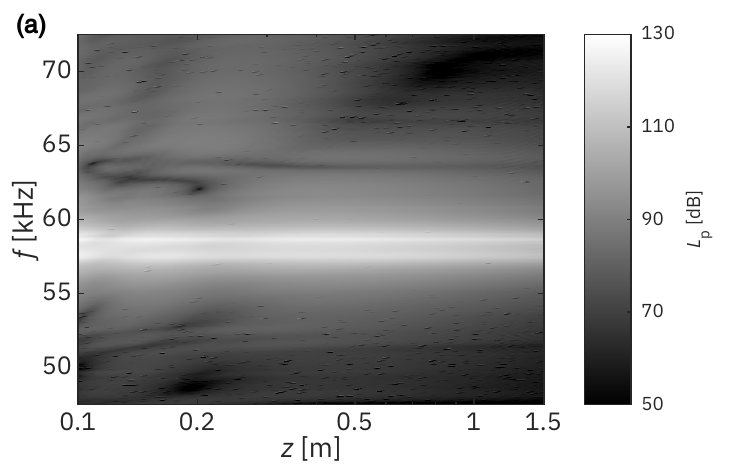}
        \caption{}
        \label{fig:ultra_pc}
    \end{subfigure}
    \begin{subfigure}[b]{8.5cm}
        \centering
        \includegraphics[width=\linewidth]{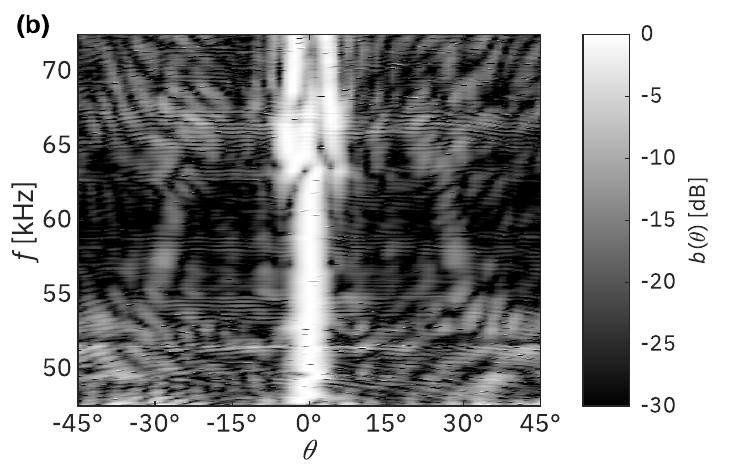}
        \caption{}
        \label{fig:ultra_bp}
    \end{subfigure}
    \\
    \begin{subfigure}[b]{4.25cm}
        \centering
        \includegraphics[width=\linewidth]{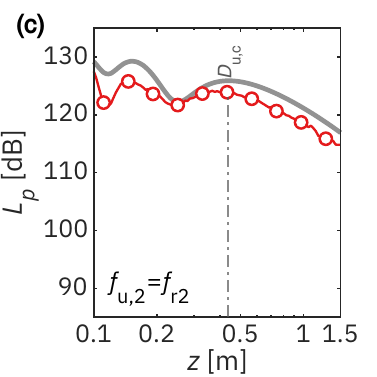}
        \caption{}
        \label{fig:ultra_pc_1}
    \end{subfigure}
    \begin{subfigure}[b]{4.25cm}
        \centering
        \includegraphics[width=\linewidth]{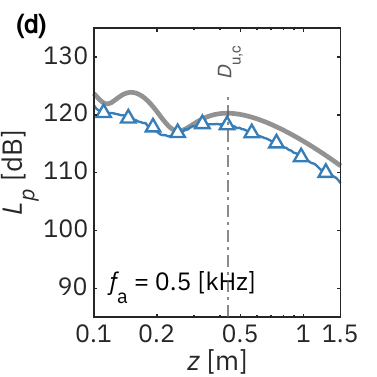}
        \caption{}
        \label{fig:ultra_pc_2}
    \end{subfigure}
    \begin{subfigure}[b]{4.25cm}
        \centering
        \includegraphics[width=\linewidth]{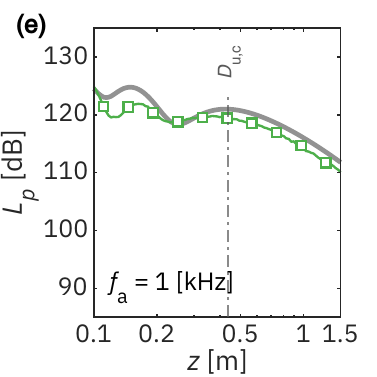}
        \caption{}
        \label{fig:ultra_pc_3}
    \end{subfigure}
    \begin{subfigure}[b]{4.25cm}
        \centering
        \includegraphics[width=\linewidth]{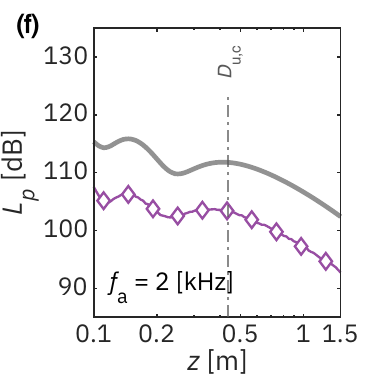}
        \caption{}
        \label{fig:ultra_pc_4}
    \end{subfigure}
    \\
    \begin{subfigure}[b]{4.25cm}
        \centering
        \includegraphics[width=\linewidth]{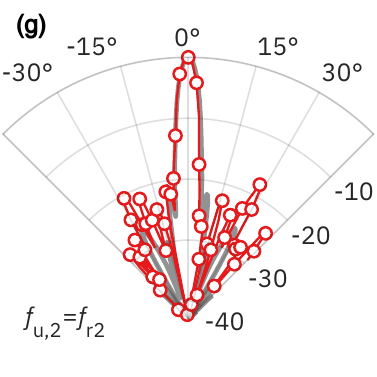}
        \caption{}
        \label{fig:ultra_bp_1}
    \end{subfigure}
    \begin{subfigure}[b]{4.25cm}
        \centering
        \includegraphics[width=\linewidth]{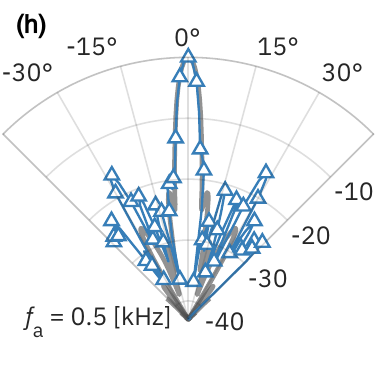}
        \caption{}
        \label{fig:ultra_bp_2}
    \end{subfigure}
    \begin{subfigure}[b]{4.25cm}
        \centering
        \includegraphics[width=\linewidth]{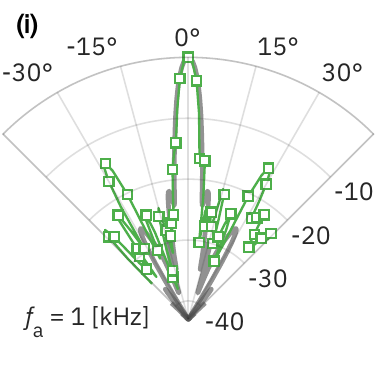}
        \caption{}
        \label{fig:ultra_bp_3}
    \end{subfigure}
    \begin{subfigure}[b]{4.25cm}
        \centering
        \includegraphics[width=\linewidth]{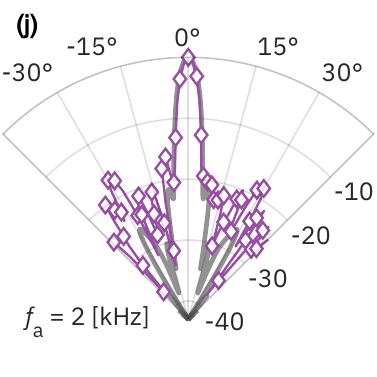}
        \caption{}
        \label{fig:ultra_bp_4}
    \end{subfigure}
    \caption{
        \label{fig:ultra}
        Ultrasound propagation characteristics and BPs generated by the SPPAL under a driving voltage of $5$~V. (a) Acoustic pressure field plotted as a function of frequency $f$ and propagation distance $z$. (b) BP of the ultrasound field as a function of frequency $f$ and angle $\theta$. (c), (g) PC and BP at the carrier resonance frequency $f_\mathrm{u,2} = f_\mathrm{r2}$. (d)-(f), (h)-(j) PCs and BPs at the LSB frequencies $f_\mathrm{u,1}$ corresponding to audio modulation frequencies $f_\mathrm{a} = 0.5$, $1$, and $2$~kHz, where $f_\mathrm{u,1} = f_\mathrm{u,2} - f_\mathrm{a}$. Markers indicate experimental measurements, while thick gray solid lines denote theoretical predictions based on Method B.
    }
\end{figure*}

Figure~\ref{fig:ultra_pc} illustrates the ultrasound propagation characteristics of the SPPAL driven at $5$~V, showing the acoustic pressure field as a function of distance $z$ and frequency $f$. Two distinct resonances appear near $60$~kHz. While appreciable pressure levels are also observed in the USB range, they are not practically usable due to unintended transitions to the 9th vibrational mode, as discussed in Sec.~\ref{subsubsec:422} and further detailed in the next section. Figures~\ref{fig:ultra_pc_1}-\ref{fig:ultra_pc_4} compare the experimentally measured PCs (markers) with theoretical predictions from Method B (thick gray solid lines), evaluated at the carrier resonance $f_\mathrm{u,2} = f_\mathrm{r2}$ and corresponding LSB frequencies $f_\mathrm{u,1} = f_\mathrm{u,2} - f_\mathrm{a}$ for $f_\mathrm{a} = 0.5$, $1$, $2$~kHz. The experimental results closely match the analytical predictions, with clear acoustic peaks near the design target $\mathscr{D}_\mathrm{u,c}$. These results validate that the SPPAL emulates the radiation behavior of a RP, as established in Sec.~\ref{subsec:22}, and support the use of the ER in the multi-objective optimization framework by confirming the predictive reliability of the analytical approach.

\paragraph{\label{paragraph:4313} Beam pattern}

Figure~\ref{fig:ultra_bp} presents the BP of the ultrasound generated by the SPPAL at a driving voltage of $5$~V, showing acoustic pressure as a function of frequency $f$ and angle $\theta$. Below approximately $63$~kHz, the beam exhibits a well-focused main lobe along the acoustic axis. However, near $63$~kHz, a transition to the 9th vibrational mode occurs, degrading beam directivity due to lobe splitting and loss of coherence. This limits the use of USB frequencies for audio modulation, as discussed in Sec.~\ref{subsubsec:442}. Figures~\ref{fig:ultra_bp_1}-\ref{fig:ultra_bp_4} compare measured BPs (markers) with theoretical predictions from Method B (thick gray solid lines), evaluated at $f_\mathrm{a} = 0.5$, $1$, $2$~kHz.

\subsubsection{\label{subsubsec:432} Audio sound}

This section presents the measurement of audio sound fields generated by the SPPAL. Overall, the experimental results show good agreement with the theoretical pr edictions across the audio frequency range. However, at $2$~kHz, a noticeable deviation is observed between the measured and predicted values. The underlying causes of this discrepancy are discussed in the Sec.~\ref{subsec:44}.

\paragraph{\label{paragraph:4321} Frequency response}

\begin{figure}[ht]
    \centering
    \includegraphics[width=\linewidth]{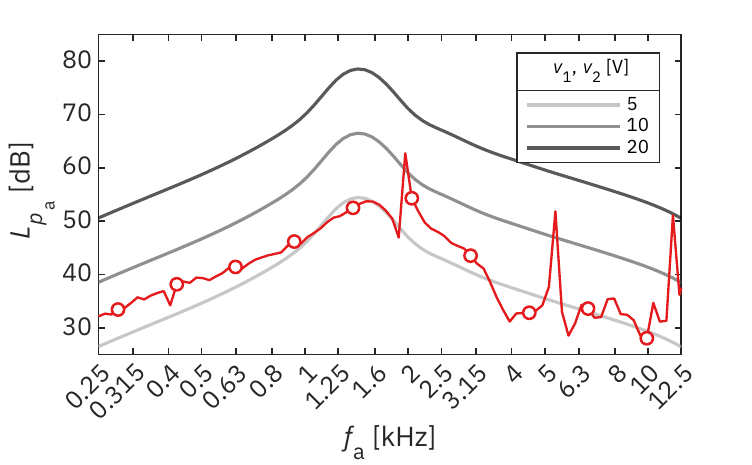}
    \caption{
        \label{fig:audio_fr}
        Measured audio frequency response of the SPPAL driven by LSB-AM at $\mathscr{D}_\mathrm{a,c} = 0.5$~m. The carrier and sideband signals were both driven at $10$~V. Markers denote experimental data, and gray lines indicate theoretical predictions based on Methods B for different driving voltages ($5$, $10$, and $20$~V).
    }
\end{figure}

Figure~\ref{fig:audio_fr} illustrates the measured audio frequency response of the SPPAL driven by LSB-AM, recorded at the $\mathscr{D}_\mathrm{a,c} = 0.5$~m. Both the carrier and sideband signals were driven at $10$~V. The experimental results (markers) reveal an enhancement in the audio output near the frequency region associated with the DR spacing, aligning with the intended design. The thick gray solid lines represent analytical predictions computed using Methods B for driving voltages of $5$, $10$, $20$~V. As previously observed in Figs.~\ref{fig:vib_fr} and \ref{fig:ultra_fr}, deviations between the fabricated transducer and its target specifications result in some discrepancy between experimental and theoretical curves. Nevertheless, the measured trends are consistent with simulation results.

\paragraph{\label{paragraph:4322} Propagation curve}

\begin{figure*}[ht]
    \begin{subfigure}[b]{4.25cm}
        \centering
        \includegraphics[width=\linewidth]{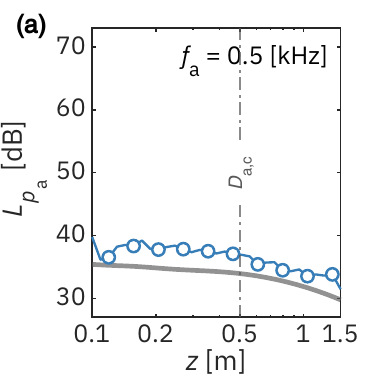}
        \caption{}
        \label{fig:audio_pc_1}
    \end{subfigure}
    \begin{subfigure}[b]{4.25cm}
        \centering
        \includegraphics[width=\linewidth]{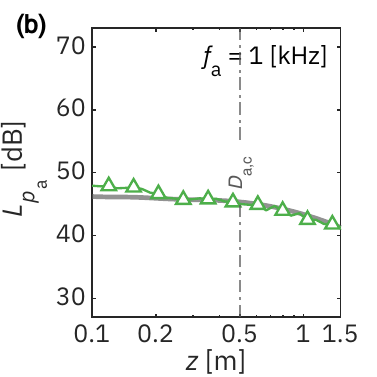}
        \caption{}
        \label{fig:audio_pc_2}
    \end{subfigure}
    \begin{subfigure}[b]{4.25cm}
        \centering
        \includegraphics[width=\linewidth]{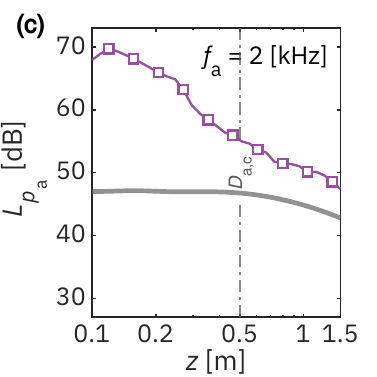}
        \caption{}
        \label{fig:audio_pc_3}
    \end{subfigure}
    \begin{subfigure}[b]{4.25cm}
        \centering
        \includegraphics[width=\linewidth]{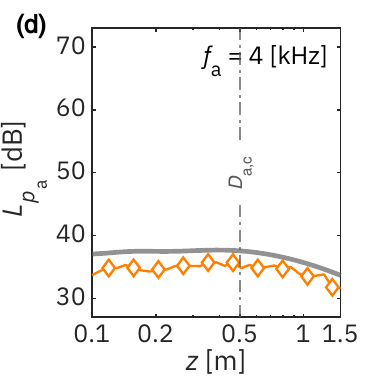}
        \caption{}
        \label{fig:audio_pc_4}
    \end{subfigure}
    \begin{subfigure}[b]{4.25cm}
        \centering
        \includegraphics[width=\linewidth]{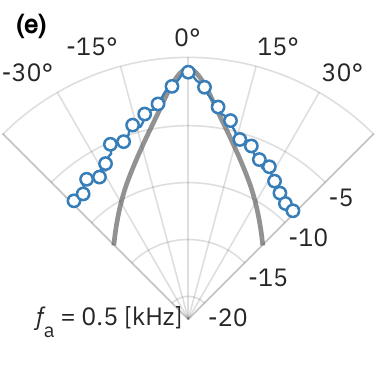}
        \caption{}
        \label{fig:audio_bp_1}
    \end{subfigure}
    \begin{subfigure}[b]{4.25cm}
        \centering
        \includegraphics[width=\linewidth]{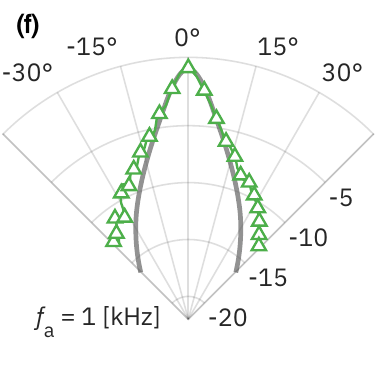}
        \caption{}
        \label{fig:audio_bp_2}
    \end{subfigure}
    \begin{subfigure}[b]{4.25cm}
        \centering
        \includegraphics[width=\linewidth]{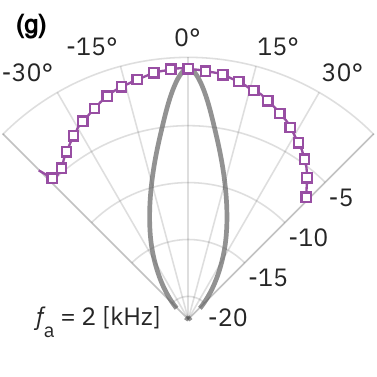}
        \caption{}
        \label{fig:audio_bp_3}
    \end{subfigure}
    \begin{subfigure}[b]{4.25cm}
        \centering
        \includegraphics[width=\linewidth]{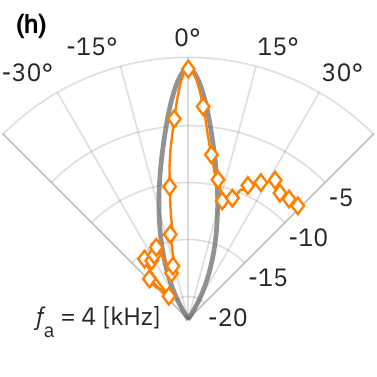}
        \caption{}
        \label{fig:audio_bp_4}
    \end{subfigure}
    \caption{
        \label{fig:audio}
        PCs and BPs of audio frequencies generated by the SPPAL under LSB-AM excitation at a driving voltage of $10$~V. (a)-(d) PCs for audio frequencies $f_\mathrm{a} = 0.5$, $1$, $2$, and $4$~kHz, respectively. (e)-(h) Corresponding BPs measured at the audio CD $\mathscr{D}_\mathrm{a,c} = 0.5$~m. Experimental results are shown as colored markers; theoretical predictions (Method B) are shown as solid gray lines. The BP at $2$~kHz exhibits noticeable deviation from the theoretical prediction.
    }
\end{figure*}

Figures~\ref{fig:audio_pc_1}-\ref{fig:audio_pc_4} illustrates the PCs of audio frequencies $f_\mathrm{a} = 0.5, 1, 2,4$~kHz generated by the SPPAL driven at $10$~V. As designed, each curve exhibits a peak at the $\mathscr{D}_\mathrm{a,c}=0.5$~m. This experimental validation confirms the appropriateness of employing a RP as the fundamental design reference for the SPPAL, as initially proposed in Sec.~\ref{subsec:21}. The audio CD thus emerges as an optimal target position for listeners due to the maximized cumulative effect of the audio signals.

\paragraph{\label{paragraph:4323} Beam pattern}

Figures~\ref{fig:audio_bp_1}-\ref{fig:audio_bp_4} presents the measured BPs of the audio frequencies $f_\mathrm{a} = 0.5, 1, 2, 4$~kHz generated by the SPPAL, recorded at $\mathscr{D}_\mathrm{a,c} = 0.5$~m. The experimental results (colored markers) show good agreement with analytical predictions (solid gray lines), also with the notable exception of the BP at $f_\mathrm{a} = 2$~kHz, where considerable deviation is observed.

\subsection{\label{subsec:44} Combination resonance}

\begin{figure*}[ht]
    \begin{subfigure}[b]{8.5cm}
        \centering
        \includegraphics[width=\linewidth]{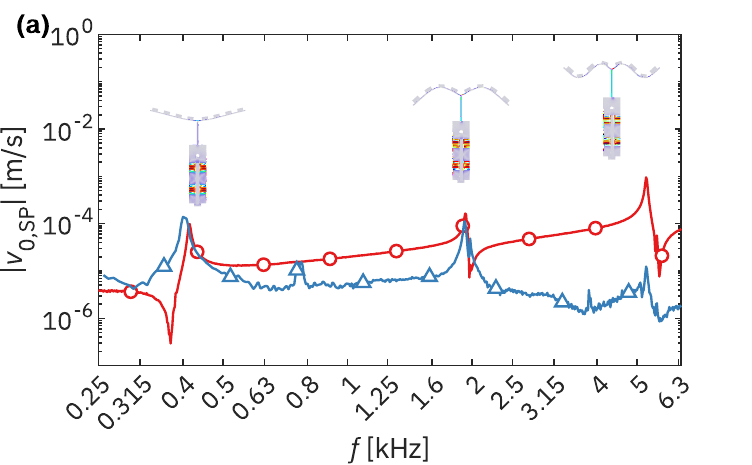}
        \caption{}
        \label{fig:combi_vfr}
    \end{subfigure}
    \begin{subfigure}[b]{8.5cm}
        \centering
        \includegraphics[width=\linewidth]{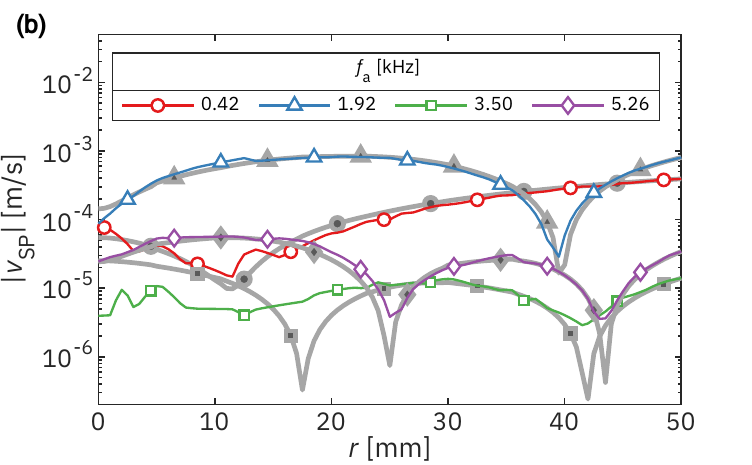}
        \caption{}
        \label{fig:combi_ma}
    \end{subfigure}
    \\
    \begin{subfigure}[b]{4.25cm}
        \centering
        \includegraphics[width=\linewidth]{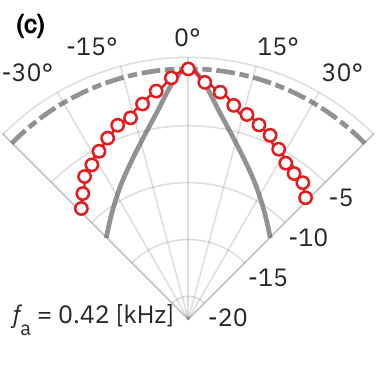}
        \caption{}
        \label{fig:combi_bp_1}
    \end{subfigure}
    \begin{subfigure}[b]{4.25cm}
        \centering
        \includegraphics[width=\linewidth]{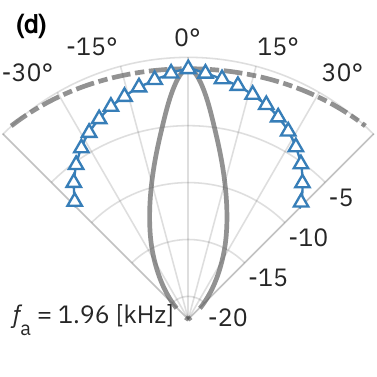}
        \caption{}
        \label{fig:combi_bp_2}
    \end{subfigure}
    \begin{subfigure}[b]{4.25cm}
        \centering
        \includegraphics[width=\linewidth]{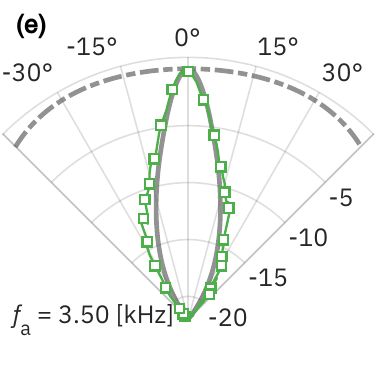}
        \caption{}
        \label{fig:combi_bp_3}
    \end{subfigure}
    \begin{subfigure}[b]{4.25cm}
        \centering
        \includegraphics[width=\linewidth]{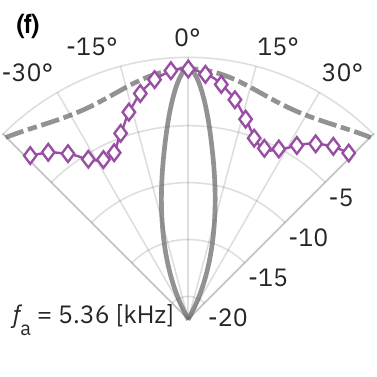}
        \caption{}
        \label{fig:combi_bp_4}
    \end{subfigure}
    \caption{
        \label{fig:combi}
        Experimental characterization of combination resonance in a SPPAL. (a) Velocity frequency response at the plate center under linear ($\circ$) and LSB-AM ($\triangle$) excitation. Resonance peaks corresponding to CR are identified, with overlaid scaled FEM mode shapes indicating the associated structural modes. (b) Experimental modal shapes measured at CR frequencies, showing close agreement with FEM-predicted linear modes (gray lines). (c)-(f) Measured beam patterns at CR frequencies, compared against Method B (gray solid lines) and FEM-based linear predictions (gray dash-dotted lines).
    }
\end{figure*}

Combination resonance (CR)\cite{Nayfeh1995, Yamamoto1964} refers to a nonlinear phenomenon in which intermodulation components excite the natural frequencies of the SPPAL, resulting in unintended mechanical responses. In the context of a SPPAL, this occurs when the difference-frequency components generated by LSB-AM coincide with the plate's inherent vibrational modes. When such resonance is triggered, the plate emits low-frequency audio sound directly through its structural vibration, rather than via nonlinear interaction in air. The amplitude of this direct audio radiation often exceeds that of the audio sound produced, meaning the perceived audio field becomes dominated by structural vibration rather than the intended airborne nonlinear process. Consequently, the presence of CR highlights the need for further investigation into its suppression or compensation to fully harness the potential of the SPPAL in directional audio applications.

\subsubsection{\label{subsubsec:441} Frequency response of plate's center velocity}

Figure~\ref{fig:combi_vfr} illustrates the velocity frequency responses measured at the center of the SPPAL plate in the audio frequency range. The circular markers represent the linear response under $5$~V excitation, while the triangular markers correspond to the response under LSB-AM excitation with both carrier and sideband signals driven at $5$~V. Distinct peaks appear near $0.4$, $2$, and $5$~kHz in both cases, indicating that the plate's structural resonances are excited by intermodulation distortion (IMD). Although the spectrum analyzer guarantees low IMD, the LDV's IMD characteristics are unspecified. Thus, the velocity magnitudes under LSB-AM may not be quantitatively reliable. Nonetheless, the consistent appearance of peaks at the known resonance frequencies supports the occurrence of CRs.

\subsubsection{\label{subsubsec:442} Experimental modal analysis}

Figure~\ref{fig:combi_ma} presents the measured modal shapes of the SPPAL plate, acquired using LDV at frequencies where CRs were observed. For comparison, the thick gray lines show the corresponding linear modal shapes obtained from FEM simulations, amplitude-scaled to match the experimental data. The measured shapes under LSB-AM excitation closely align with the predicted linear mode shapes, confirming that the excited structural responses stem from the plate's inherent resonances. As a reference, the modal shape at $3.5$~kHz is included as a control case, representing a frequency at which CR does not occur.

\subsubsection{\label{subsubsec:443} Beam pattern}

Figures~\ref{fig:combi_bp_1}-\ref{fig:combi_bp_4} show the BPs measured at frequencies where CRs occur. The thick gray solid lines represent analytical predictions obtained using Method B, while the dash-dotted lines correspond to FEM simulations under linear excitation at the same frequencies. At $3.5$~kHz, the measured BP closely matches the analytical prediction. In contrast, at CR-affected frequencies, the measured patterns exhibit significant deviations from both predictions. These results suggest that under CR conditions, the audio sound field generated by the SPPAL becomes highly unpredictable and cannot be reliably described by linear models.

\subsection{\label{subsec:45} Experiment setup}

\begin{figure}[ht]
    \begin{subfigure}[b]{8.5cm}
        \centering
        \includegraphics[width=\linewidth]{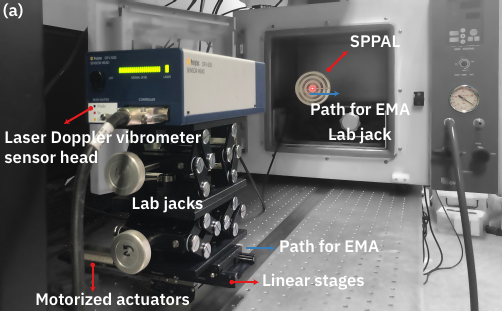}
        \caption{}
        \label{fig:vib_exp}
    \end{subfigure}
    \begin{subfigure}[b]{8.5cm}
        \centering
        \includegraphics[width=\linewidth]{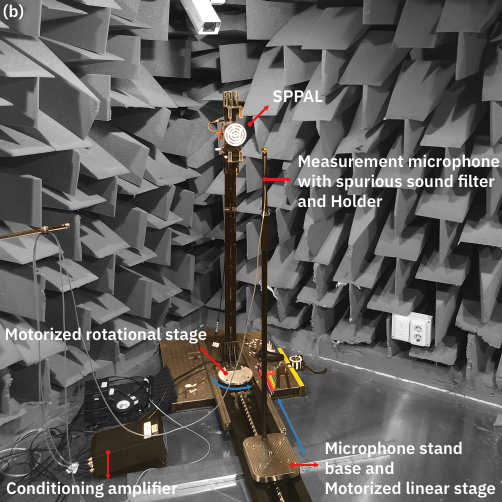}
        \caption{}
        \label{fig:aco_exp}
    \end{subfigure}
    \caption{
        \label{fig:exp}
        Experimental setup for measuring the vibrational and acoustic characteristics of the SPPAL. (a) Measurement of center velocity and radial mode shape using a LDV traversing along the plate radius. (b) Acoustic field measurement in a semi-anechoic chamber.
    }
\end{figure}

\subsubsection{\label{subsubsec:451} Vibrational characteristics}

To measure the plate's center velocity and radial mode shape of the SPPAL, the device was mounted on a lab jack using a custom-designed jig. The vibration of the plate was measured using a single-point laser Doppler vibrometer (LDV; OFV-505, OFV-5000, VD-09, Polytec), while the LDV sensor head traversed the radial axis of the plate using a pair of motorized linear stages (LTA-HS, Newport) connected in series. The actuators of the linear stages were controlled by a motion controller (ESP301, Newport). The transducer was driven by a dynamic signal analyzer (SR785, Stanford Research Systems) through a power amplifier (HSA4052, NF Corporation), and the output signal from the LDV was recorded simultaneously. All measurement instruments were controlled and synchronized using MATLAB 2024b.

\subsubsection{\label{subsubsec:433} Acoustic characteristics}

Acoustic measurements were conducted in a semi-anechoic chamber at Pohang University of Science and Technology. The chamber's background noise level, free-field effective volume, absorption coefficient, and low cutoff frequency were below $30$~dB SPL, $3\times3\times2\mathrm{m}^3$, $0.99$, and $150$~Hz, respectively. The temperature and relative humidity during the measurements were maintained at $20$\textcelsius\ and $70$\% RH. The SPPAL was mounted on a motorized rotational stage using a custom jig, positioned $1.4$~m above the floor. Acoustic fields were measured using Brüel \& Kjær 1/8" pressure-field microphone type 4138 for ultrasound and 1/2" microphone type 4192 for audio sound. The microphones, mounted on a motorized linear stage via a custom holder, were interchangeable for each measurement. For both ultrasound and audio measurements, the microphone was oriented at a $90$\textdegree\ angle of incidence to enable free-field correction and mitigate spurious sound in complex near-field measurements.\cite{Zhong2024,Ju2010,Nomura2022} Additionally, a spurious sound filter topologically equivalent to a half-wavelength resonator was applied to the microphone assembly to further suppress spurious components.\cite{Kim2025} A dynamic signal analyzer (SR785, Stanford Research Systems) drove the transducer through a power amplifier (HSA4052, NF Corporation), and the measurement signals were acquired using a conditioning amplifier type 2690. MATLAB 2024b was employed for data acquisition and instrument control.

\section{\label{sec:5} Conclusion}

This study introduced and comprehensively examined SPPAL, demonstrating its potential as an effective alternative to conventional transducer array-based PAL systems. The investigation encompassed acoustic modeling, transducer design optimization, and experimental validation. The key contribution lies in quantitatively establishing an equivalence between the stepped plate (SP) and a baffled rigid piston (RP) using the equivalence ratio (ER), enabling efficient and efficient preliminary design approximations; this, in turn, facilitated a broad parametric study.
\par
To address the intrinsic limitations of low-frequency sound pressure common in conventional PAL systems, a dual-resonance (DR) approach was proposed and validated. This DR strategy employed multi-objective optimization across a comprehensive set of design parameters. Utilizing an Approximated 3D modeling method, this study effectively balanced computational efficiency and model accuracy, significantly extending feasible design parameter ranges beyond traditional quarter-wavelength constraints up to half-wavelength diameters. This advancement allowed the practical choice of larger horn-end radii, thus directly mitigating critical manufacturing challenges. Optimal design parameters were identified by evaluating trade-offs among carrier frequency, aperture radius, and transducer geometry. Clear trends observed from the impact of these parameters on audio SPL elucidate key design parameters, providing practical guidelines for future developments in SPPAL systems or analogous flexural ultrasonic transducers.
\par
Comprehensive experimental modal analyses provided robust validation of the theoretical and computational results, distinctly revealing the modal behavior of the stepped plate and explicitly confirming the effectiveness of the annular steps in phase compensation. Acoustic measurements reinforced these findings, affirming the accuracy of the proposed modeling approaches and the validity of the design methodology. Additionally, the experimental characterization identified the presence of combination resonance (CR), a structural excitation resulting from intermodulation phenomena. CR was experimentally demonstrated as inherent to the SPPAL due to its single-transducer configuration combined with high-order flexural mode operation of an extended radiating surface. Such resonance is not typically observed in conventional PAL systems employing arrays of single-resonance transducers. Although CR may induce unintended structural audio emissions, its identification highlights an essential area for future investigation into resonance mitigation strategies for more robust and reliable SPPAL.
\par
Overall, this work establishes a comprehensive workflow for the design and analysis of SPPAL--from theoretical modeling to practical implementation. Beyond demonstrating the effectiveness of the proposed design scheme, the study also identifies the inherent limitations of the SPPAL concept, particularly those arising from the structural dynamics of high-order-mode operation. These findings are expected to inform the design of future flexural transducer systems, especially those adopting similar principles. Moreover, the proposed design scheme offers a valuable foundation for the development of high-performance PAL transducers and related nonlinear acoustic devices.

\begin{acknowledgments}
This work was supported by the Industrial Strategic Technology Development Program (Korea-led for K-sensor technology for market leadership) (RS-2022-00154770, Next-generation piezoelectric based motion sensor platform) funded By the Ministry of Trade Industry \& Energy (MOTIE, Korea).
\end{acknowledgments}

\section*{Author Declarations}
The authors have no conflicts to disclose.

\section*{Data Availability}
The data that support the findings of this study are available from the corresponding author upon reasonable request.

\bibliography{Notes_arXiv}

\end{document}